\documentclass{jfm}
\usepackage{graphicx}
\usepackage{epstopdf, epsfig}
\usepackage{caption}
\usepackage{color, soul}
\usepackage[colorinlistoftodos]{todonotes}

\usepackage{longtable} 
\usepackage{multirow} 
\usepackage{amsmath} 
\usepackage[T1]{fontenc} 
\usepackage{tabularx} 
\usepackage{changepage} 
\usepackage{natbib} 
\usepackage{float} 
\usepackage[running]{lineno} 
\usepackage{xcolor} 
\newcommand{\var}[1]{{\operatorname{#1}}} 
\usepackage{relsize} 
\usepackage{url} 
\shorttitle{The role of intermittent heat transport towards Reynolds stress anisotropy}
\shortauthor{Subharthi Chowdhuri, Siddharth Kumar and Tirtha Banerjee}
\title{Revisiting the role of intermittent heat transport towards Reynolds stress anisotropy in convective turbulence}

\author{Subharthi Chowdhuri\aff{1}
  \corresp{\email{subharthi.cat@tropmet.res.in}},
  Siddharth Kumar\aff{1}
  \and Tirtha Banerjee\aff{2}}

\affiliation{\aff{1}Indian Institute of Tropical Meteorology, Ministry of Earth Sciences, India\\
Dr. Homi Bhaba Road, Pashan, Pune-411008
\aff{2}Department of Civil and Environmental Engineering, University of California, Irvine, CA 92697, USA
}

\begin{document}
\maketitle

\begin{abstract}
Thermal plumes are the energy containing eddy motions that carry heat and momentum in a convective boundary layer. The detailed understanding of their structure is of fundamental interest for a range of applications, from wall-bounded engineering flows to quantifying surface-atmosphere flux exchanges. We address the aspect of Reynolds stress anisotropy associated with the intermittent nature of heat transport in thermal plumes by performing an invariant analysis of the Reynolds stress tensor in an unstable atmospheric surface layer flow, using a field-experimental dataset. Given the intermittent and asymmetric nature of the turbulent heat flux, we formulate this problem in an event-based framework. In this approach, we provide structural descriptions of warm-updraft and cold-downdraft events and investigate the degree of isotropy of the Reynolds stress tensor within these events of different sizes. We discover that only a subset of these events are associated with the least anisotropic turbulence in highly-convective conditions. Additionally, intermittent large heat flux events are found to contribute substantially to turbulence anisotropy under unstable stratification. Moreover, we find that the sizes related to the maximum value of the degree of isotropy do not correspond to the peak positions of the heat flux distributions. This is because, the vertical velocity fluctuations pertaining to the sizes associated with the maximum heat flux, transport significant amount of streamwise momentum. A preliminary investigation shows that the sizes of the least anisotropic events probably scale with a mixed-length scale ($z^{0.5}\lambda^{0.5}$, where $z$ is the measurement height and $\lambda$ is the large-eddy length scale).
\end{abstract}

\begin{keywords}
\end{keywords}

\section{Introduction}
Taylor's statistical theory of turbulence states that the turbulence is isotropic if the average value of any function of the velocity components, defined in relation to a given set of axes, is unaltered under axis rotation \citep{taylor1935statistical}. However, the condition of isotropy is not satisfied for the energy-containing scales of turbulence, since no energy production can happen for isotropic turbulence due to its directional independence \citep{tennekes1972first,wyngaard2010turbulence}. Several metrics have been used to quantify turbulence anisotropy (see table \ref{tab:1} for a brief review) and out of those one of the metrics (also used in this study) to quantify the anisotropic signatures of the energy containing motions at a point in the flow is the anisotropy Reynolds stress tensor \citep{krogstad2000invariant,smyth2000anisotropy,antonia2001turbulence,pouransari2015statistical}. The anisotropy Reynolds stress tensor ($b_{ij}$) is defined in a Cartesian co-ordinate system as, 
\begin{equation}
b_{ij} =\frac{\overline{u^{\prime}_{i}u^{\prime}_{j}}}{2q}-\frac{1}{3}\delta_{ij}, \ q=\frac{\overline{u^{\prime}_{k}u^{\prime}_{k}}}{2}, 
\label{RS}
\end{equation}
where overbar indicates averaging over time, $u^{\prime}_{i}$ are the turbulent fluctuations in the velocity field ($i=1,2,3$), $\delta_{ij}$ is the Kronecker delta, and $q$ is the averaged turbulent kinetic energy. This tensor becomes zero in an isotropic turbulence and its anisotropy is quantified by using the invariants of $b_{ij}$, an approach pioneered by \citet{lumley1977return} and \citet{lumley1979computational}, known as invariant analysis. The invariants of the anisotropy Reynolds stress tensor have been used extensively in the the context of wall-bounded neutral flows (without the effect of buoyancy) to deduce the anisotropic characteristics of the energy-containing motions  \citep{shafi1995anisotropy,antonia2001turbulence,smalley2002reynolds,ashrafian2006structure}. 

In convective turbulence, buoyant structures, such as thermal plumes, are the energy containing motions that transport heat and drive the flow \citep{celani2001thermal,shang2003measured}. These thermal plumes are well-organized structures of warm-rising (warm-updrafts) and cold-descending (cold-downdrafts) fluid, which generate ramp-cliff patterns in temperature time series when passing a thermal probe \citep{zhou2002plume}. \citet{shang2003measured} have shown that in turbulent Rayleigh B\'{e}nard convection, the time series of the instantaneous vertical heat flux associated with the thermal plumes displays intermittent characteristics. Intermittency is defined as a property of the turbulent signal which is quiescent for much of the time and occasionally burst into life with unexpectedly high values more common than in a Gaussian signal \citep[e.g.,][]{davidson2015turbulence}. However, the effect of this intermittent heat transport on the anisotropic fluctuations in the velocity field of convective turbulence is not yet well understood, as acknowledged by \citet{pouransari2015statistical}. This problem is particularly relevant for the surface layer of a convectively driven atmospheric boundary layer, where the most prevalent coherent structures are the thermal plumes and the heat transport characteristics associated with these plumes appear to be intermittent \citep{duncan1992method,katul1994conditional,caramori1994structural,chu1996probability,katul1997turbulent_b,katul1997ejection_c}. 

The previous works on the atmospheric surface layer (ASL) plumes have focused on: (a) deducing their detailed structures and dynamics \citep{wilczak1984large,zhuang1995dynamics}; (b) identifying the coupling between the surface and air temperatures \citep{garai2011air,garai2013interaction}; and (c) investigating the difference in the Monin-Obukhov similarity functions by conditioning on the updraft and downdraft motions \citep{li2018implications,fodor2019role}. However, some early investigators noted that in an unstable ASL there were certain intermittent bursts in the upward heat flux, persisting for around 10-20 s of duration, which were associated with large downward momentum transport \citep{kaimal1969measurement,kaimal1970case,haugen1971experimental}. They commented that the vertical velocity fluctuations associated with these heat flux events could either transport momentum downward in large bursts, or transport it upward. \citet{businger1973note} coined these intermittent momentum bursts associated with the heat flux events as ``convection-induced stress''. Recently, \citet{lotfy2019characteristics} also obtained the same result from a field experiment in an unstable ASL, where they observed that the persistent warm-updrafts of 10-20 s duration were associated with large amount of momentum flux in the downward direction. By investigating the large eddy simulation results in convective conditions, \citet{salesky2018buoyancy} interpreted this phenomenon as a buoyancy-dominated scale modulation effect. They explained that under highly-convective conditions, the small-scale turbulence is excited in the updraft regions and suppressed in downdraft regions, leading to intermittent periods of small-scale excitation in the momentum fluxes.

From the discussion above, it becomes apparent that in an unstable ASL, the vertical velocity fluctuations associated with the coherent heat flux events could transport large amount of momentum in intermittent bursts, either in upward or downward direction. Since only the anisotropic part of the velocity fluctuations can carry momentum \citep{dey2018turbulent,konozsy2019new}, it indicates that the Reynolds stress anisotropy associated with these coherent heat flux events must be different from the averaged whole flow. Therefore, studying the role of intermittent heat flux events towards the anisotropy in the velocity fluctuations is of practical importance in the context  of ASL turbulence. For a systematic investigation of this problem, invariant analysis of the anisotropy Reynolds stress tensor in an event-based framework is a well-suited approach. 

The event based approach in turbulence is based on the fact that coherent physical structures exist in a turbulent flow \citep{chapman1985observations,narasimha1990turbulent,kailas1994similarity,hogstrom1996organized,baron1997turbulent,narasimha2007turbulent}. Specifically, \citet{narasimha2007turbulent} mentioned that in this approach, the turbulent field can be expressed in terms of events, given that its types, magnitudes, arrival times, etc. are defined properly. The interest in the event based description of turbulence started with the flow visualization studies of \citet{kline1967structure}, \citet{corino1969visual}, and \citet{kim1971production}. They observed that the flow near the wall of a boundary layer was organized into streaks of high- and low-momentum fluid. Subsequently, the low-momentum streaks were seen to intermittently erupt away from the wall in a chaotic process named bursting. This accounted for much of the outward vertical transport of momentum and the production of turbulent kinetic energy in the boundary layer. These burst events were detected from the point measurements of velocity components by quadrant analysis \citep{lu1973measurements}. A detailed review of different conditional sampling techniques to detect events in turbulence can be found in \citet{antonia1981conditional} and \citet{wallace2016quadrant}. The types of coherent structures whose signatures are associated with these events are reviewed in detail by \citet{cantwell1981organized}, \citet{robinson1991coherent}, and \citet{jimenez2018coherent}. 

\citet{sreenivasan1979local} first applied this event based approach to investigate the effect on turbulence anisotropy associated with the coherent structures in a heated turbulent jet. Based on the premise that the fine structures were superposed on the large structures, \citet{sreenivasan1979local} extracted the coherent ramp-cliff events in a heated turbulent jet, and then subtracted these patterns from the signal to get the superposed fluctuations. Their focus was to show that the skewness in the temperature gradient vanishes for the fine structures, thus confirming the local isotropy. Recently following the work of \citet{lozano2012three}, \citet{dong2017coherent} studied the connected regions of high-intensity momentum zones in three-dimensional simulations of homogeneous shear and channel flows and investigated the Reynolds stress anisotropy. They quantified anisotropy by the invariants of the Reynolds stress tensor within these high-intensity momentum zones along with their sizes; where the size was defined as the box-diagonal of the parallelepiped which circumscribed these connected regions. \citet{zhou2011disentangle} attempted to disentangle the role of thermal plumes on the velocity field in a Rayleigh B\'{e}nard convection, by studying separately the anisotropy in the inertial subrange of the positive and negative vertical velocity increments. They showed that the negative increments at small separations deviated from the Kolmogorov scaling, which they attributed to the presence of the coherent structures such as thermal plumes. 

\begin{table}
  \begin{center}
   \begin{adjustwidth}{-1cm}{}
\def~{\hphantom{0}}
  \begin{tabularx}{\textwidth}{lccc}
Authors  & Approach   &   Metric & Remarks\\[3pt]
      \\
      \multirow{2}{*}{\citet{chamecki2004local}} & Scale-  & Spectra and & Test of local\\
& decomposition & structure functions & isotropy hypothesis\\ 
\\

      \multirow{2}{*}{\citet{kurien2000anisotropic}} & Scale-  & SO(3) decomposition of & Anisotropy in small\\
& decomposition & structure functions & scale motions\\ 
\\

      \multirow{2}{*}{\citet{djenidi2012anisotropy}} & Time-averaged  & Reynolds stress and & Large- and \\
& statistics & dissipation tensors & small-scale anisotropy\\ 
\\

      \multirow{2}{*}{\citet{djenidi2009anisotropy}} & Time-averaged  & Taylor's anisotropy & Anisotropy in \\
& statistics & coefficient & energy-containing motions\\ 
\\

      \multirow{2}{*}{\citet{salesky2017nature}} & Time-averaged  & Vertical and horizontal & Anisotropy in \\
& statistics & velocity variance ratio & energy-containing motions\\ 
\\

      \multirow{2}{*}{\citet{liu2017scale}} & Scale- & Scale-decomposed& Scale description of\\
& decomposition & Reynolds stress tensor & anisotropy in an\\
& &  & urban surface layer\\
\\

      \multirow{2}{*}{\citet{dong2017coherent}} & Event based  & Reynolds stress& Reynolds stress anisotropy \\
& description & tensor &  associated with\\
& & & coherent structures\\
\\

      \multirow{2}{*}{\citet{zhou2011disentangle}} & Event based description   & Conditionally sampled& Anisotropy in \\
& and scale-decomposition & structure functions & positive and negative\\
&  &  & velocity increments\\
\\

  \end{tabularx}
  \caption{A brief summary of different approaches and metrics used to study anisotropy in a turbulent flow.}
  \label{tab:1}
   \end{adjustwidth}
  \end{center}
\end{table}

The anisotropy directly associated with the intermittent occurrences of the coherent structures is regarded as a state-of-the-art theoretical and experimental problem \citep{pouransari2015statistical}. To the best of our knowledge, very few studies have addressed this problem by adopting an event-based approach. This is particularly pertinent in the context of ASL turbulence, where there are no comprehensive studies to quantify anisotropy concomitant with the intermittent heat flux events in convective conditions. The present study attempts to fill this gap, using a field-experimental dataset. Therefore, we define our objectives as:
\begin{enumerate}
\item To investigate the detailed correspondence between the heat flux events and turbulence anisotropy in an unstable ASL.
\item To formulate a structural description of the heat flux events and investigate whether they have any characteristic length scales associated with least anisotropic turbulence.
\end{enumerate}    

The present paper is organized in three different sections. In \S\ref{sec:data} we describe the dataset and methodology to develop various statistical measures to quantify anisotropy associated with the heat flux events. In \S\ref{results} we present and discuss the results and in \S\ref{conclusion} we conclude our findings and provide future directions for further research. 

\section{Data and Methodology}\label{sec:data}
We have used the dataset from Surface Layer Turbulence and Environmental Science Test (SLTEST) experiment. The SLTEST experiment was conducted over a flat and homogeneous terrain at the Great Salt Lake desert in Utah, USA (40.14$^\circ$ N, 113.5$^\circ$ W), with the aerodynamic roughness length ($z_0$) being $z_{0}\approx$ 5 mm \citep{metzger2007near}. The SLTEST site characteristics and the high quality of the dataset have been documented in details in many previous studies \citep{hutchins2007evidence,hutchins2012towards,chauhan2013structure,marusic2013logarithmic}. In this experiment, nine north-facing sonic anemometers (CSAT3, Campbell Scientific, Logan, USA) were installed on a 30-m tower approximately logarithmically at $z=$ 1.4, 2.1, 3, 4.3, 6.1, 8.7, 12.5, 17.9, 25.7 m, levelled to within $\pm$0.5$^\circ$ from the true vertical. All CSAT3 sonic anemometers were synchronized in time and the sampling frequency was set at 20 Hz. The experiment ran continuously for nine days from 26 May 2005 to 03 June 2005.  

\subsection{Data Processing}
\label{data_process}
The data were divided into 30-min periods containing the 20-Hz measurements of the three wind components and the sonic temperature from all the nine sonic anemometers. To select the 30-min periods for analysis, we followed these standard procedures listed below:
\begin{enumerate}
	\item The 30-min periods were selected from the fair weather conditions during the daytime periods with no rain.
	\item The time series of all the three components of velocity and sonic temperature were plotted and visually checked. No electronic spikes were found in the data \citep{vickers1997quality}.
	\item The horizontal wind direction sector was limited to $-30^\circ < \theta < 30^\circ$ (where $\theta$ is the horizontal wind direction from the North).
	\item The coordinate systems of all the nine sonic anemometers were rotated in the streamwise direction by applying the double-rotation method of \citet{kaimal1994atmospheric} for each 30-min period. The turbulent fluctuations in the wind components ($u^{\prime}$, $v^{\prime}$, and $w^{\prime}$ in the streamwise, cross-stream, and vertical directions respectively), and in the sonic temperature ($T^{\prime}$) were calculated after removing the 30-min linear trend from the associated variables \citep{donateo2017case}.
    \item Only those 30-min periods were chosen when the surface layer was unstable, i.e. the sensible heat flux was positive at all the nine measurement heights, and the vertical variations in the 30-min averaged momentum and heat fluxes were less than 10 \%.
	\end{enumerate}

Application of all these checks resulted in a total of 29 periods suitable for our analysis. For these periods $\sigma_{u}/\overline{u}$ was less than 0.2, so the Taylor's hypothesis could be assumed to be valid \citep{willis1976use}. The Obukhov length ($L$) was calculated for each of these 30-min periods as,
\begin{equation}
L = -\frac{u_*^3T_0}{kgH_0},
\label{obukhov}
\end{equation}
where $T_{0}$ is the surface air temperature, computed from the mean sonic temperature at $z=$ 1.4 m, $g$ is the acceleration due to gravity (9.8 m s$^{-2}$), $H_{0}$ is the surface kinematic heat flux, computed as $\overline{w^{\prime}T^{\prime}}$ at $z=$ 1.4 m (by constant flux layer assumption), $k$ is the von K\'arm\'an constant (0.4), and $u_{*}$ is the friction velocity computed as,
\begin{equation}
u_{*}={({\overline{u^{\prime}w^{\prime}}}^{2}+{\overline{v^{\prime}w^{\prime}}}^{2})}^{\frac{1}{4}},
\label{fric_velo}
\end{equation}
where $\overline{u^{\prime}w^{\prime}}$ and $\overline{v^{\prime}w^{\prime}}$ are the streamwise and cross-stream momentum fluxes respectively, computed at $z=$ 1.4 m. 

The range of $-L$ values was between 2 to 20 m for these 29 periods suitable for our analysis. Since each 30-min period consisted of the nine level time-synchronized turbulence measurements from the CSAT3 sonic anemometers, a total of 261 combinations of the stability ratios ($\zeta=z/L$) were possible for these selected periods. The entire range of $-\zeta$ (12 $\leq \zeta \leq$ 0.07) was divided into six stability classes \citep{liu2011probability} and these were considered for the detailed analysis of the Reynolds stress anisotropy associated with the heat flux events (see table \ref{tab:2}). We discuss the analysis methods in the following sections. 

\begin{table}
  \begin{center}
\def~{\hphantom{0}}
  \begin{tabular}{lcc}
      \multirow{2}{*}{Stability class}  & Number of & Heights \\[3pt]
      & 30-min runs &\\
      \\
$-\zeta \ >$ 2 & 55 & $z=$ 6.1, 8.7, 12.5, 17.9, 25.7 m\\
\\
1 $< \ -\zeta \ <$ 2 & 53 & $z=$ 3, 4.3, 6.1, 8.7, 12.5, 17.9, 25.7 m\\
\\
0.6 $< \ -\zeta \ <$ 1 & 41 & $z=$ 2.1, 3, 4.3, 6.1, 8.7, 12.5, 17.9 m\\
\\
0.4 $< \ -\zeta \ <$ 0.6 & 34 & $z=$ 1.4, 2.1, 3, 4.3, 6.1, 8.7 m\\
\\
0.2 $< \ -\zeta \ <$ 0.4 & 44 & $z=$ 1.4, 2.1, 3, 4.3, 6.1 m\\
\\
0 $< \ -\zeta \ <$ 0.2 & 34 & $z=$ 1.4, 2.1, 3 m\\
  \end{tabular}
  \caption{The six different stability classes formed from the ratio $-\zeta=z/L$ in an unstable ASL flow, from highly-convective ($-\zeta>2$) to near-neutral ($0<-\zeta<0.2$). The associated heights with each of the stability classes are also given.}
  \label{tab:2}
  \end{center}
\end{table}

\subsection{Quadrant Analysis}
\label{QA}
The quadrant analysis is a conditional-sampling method of investigating the contributions to the turbulent transport of scalars and momentum in terms of the organized eddy motions present in the flow \citep{wallace2016quadrant}. The four different quadrants of the $u^{\prime}$-$w^{\prime}$ and $T^{\prime}$-$w^{\prime}$ planes are defined in table \ref{tab:3}. In the $T^{\prime}$-$w^{\prime}$ ($u^{\prime}$-$w^{\prime}$) quadrant plane, the warm-updrafts (I) (ejections (II)) and cold-downdrafts (III) (sweeps (IV)) are the down-gradient motions. On the other hand, the remaining two quadrants represent the counter-gradient motions generated due to the turbulent swirls in the flow \citep{gasteuil2007lagrangian}. 
\begin{table}
  \begin{center}
\def~{\hphantom{0}}
  \begin{tabular}{lccc}
$u^{\prime}$-$w^{\prime}$ quadrant & Quadrant name & $T^{\prime}$-$w^{\prime}$ quadrant & Quadrant name 
\\
\\
$u^{\prime}<$ 0, $w^{\prime}>$ 0 (II) & Ejection & $w^{\prime}>$ 0, $T^{\prime}>$ 0 (I) & Warm-updraft\\
$u^{\prime}>$ 0, $w^{\prime}<$ 0 (IV) & Sweep & $w^{\prime}<$ 0, $T^{\prime}<$ 0 (III) & Cold-downdraft\\
$u^{\prime}>$ 0, $w^{\prime}>$ 0 (I) & Outward-interaction & $w^{\prime}>$ 0, $T^{\prime}<$ 0 (II) & Cold-updraft\\
$u^{\prime}<$ 0, $w^{\prime}<$ 0 (III) & Inward-interaction & $w^{\prime}<$ 0, $T^{\prime}>$ 0 (IV) & Warm-downdraft\\
  \end{tabular}
  \caption{The four quadrants of $u^{\prime}$-$w^{\prime}$ and $T^{\prime}$-$w^{\prime}$ in an unstable ASL.}
  \label{tab:3}
  \end{center}
\end{table}

In the quadrant analysis method applied to the ASL, the momentum or heat flux fractions and time fractions from each quadrant of $u^{\prime}$-$w^{\prime}$ or $T^{\prime}$-$w^{\prime}$ are reported over smooth and rough surfaces \citep{mcbean1974turbulent,antonia1977similarity,narasimha2007turbulent,zou2017impact}. The flux fractions ($F_{f}$) and time fractions ($T_{f}$) for each quadrant (X) are evaluated as,
\begin{align}
\begin{split}
(F_{f})_{\rm X}=\frac{\mathlarger{\sum{\Big[(w^{\prime}x^{\prime}})I_{\rm X}\Big]}}{\sum{w^{\prime}x^{\prime}}}, \ (x=u,T)
\\
(T_{f})_{\rm X}=\frac{\sum{I_{\rm X}}}{N}, \ (\rm X=I,II,III,IV)
\end{split}
\label{fraction}
\end{align}
where,
\[
                I_{\rm X} = \begin{cases}
                                        1 &\text{if $\{w^{\prime},x^{\prime}\} \in \rm{X}$}\\
                                        0 &\text{otherwise}
                                \end{cases}
\]
and $N$ is the total number of points in a run. 

However, following \citet{chowdhuri2019representation}, we extend the quadrant analysis method to study the anisotropy Reynolds stress tensor in relation to the heat flux events occurring in $T^{\prime}$-$w^{\prime}$ quadrant plane. We normalize $w^{\prime}$ and $T^{\prime}$ by their respective standard deviations, and use the symbol $\hat{x}$ to denote the turbulent fluctuations in $x$ normalized by its standard deviation ($\hat{x}=x^{\prime}/\sigma_{x}$, where $x$ can be $u$, $w$, or $T$). Before describing the methodology, we give a short description of the anisotropy Reynolds stress tensor.

\subsubsection{Anisotropy Reynolds stress tensor}
\label{theory}
The anisotropy Reynolds stress tensor is widely used to express the anisotropy in the energy-containing motions \citep{pope2001turbulent}, and is defined in the Cartesian tensor notation as:
\begin{equation}
b_{ij} =\frac{\overline{u^{\prime}_{i}u^{\prime}_{j}}}{2q}-\frac{1}{3}\delta_{ij}, \ q=\frac{\overline{u^{\prime}_{k}u^{\prime}_{k}}}{2}, 
\label{reynolds}  
\end{equation}
where $i=$ 1, 2, and 3 denote the streamwise, cross-stream, and vertical directions, $q$ is the turbulent kinetic energy, and $\delta_{ij}$ is the Kronecker delta. Note that $b_{ij}$ is a symmetric and trace-less tensor, bounded between $-1/3 \leq b_{ij} \leq 2/3$, and equal to zero for isotropic turbulence \citep{konozsy2019new}. From the Cayley-Hamilton theorem \citep{lumley1979computational,pope2001turbulent}, the two invariants $\xi$ and $\eta$ of $b_{ij}$ are defined as,
\begin{equation}
6\xi^{3} = b_{ij}b_{jk}b_{ki},
\label{xi}
\end{equation}
and
\begin{equation}
6\eta^{2} = b_{ij}b_{ji}.
\label{eta}
\end{equation}
where $\xi$ represents the topology of the anisotropy Reynolds stress tensor and $\eta$ represents the degree of isotropy. 

The different realizable anisotropic states of turbulence are defined based on the values of $\xi$ and $\eta$ and are represented on the $\xi$-$\eta$ plane, known as the anisotropy invariant map \citep{choi2001return}. The anisotropy Reynolds stress tensor ($b_{ij}$) has three limiting anisotropic states based on the shape of the energy distribution in the three principle axes associated with the three eigenvalues and eigenvectors of $b_{ij}$, also known as the componentality of turbulence \citep{kassinos2001one,simonsen2005turbulent}. These three limiting states of $b_{ij}$ are 1-component anisotropy (rod-like energy distribution, $b_{1 \rm c}$), 2-component anisotropy (disk-like energy distribution, $b_{2 \rm c}$), and 3-component isotropy (spherical energy distribution, $b_{3 \rm c}$), represented in the principal axes coordinate system as,
\begin{equation}
b_{1 \rm c} =
  \begin{bmatrix}
    2/3 & 0 & 0 \\ \\
    0 & -1/3 & 0 \\ \\
    0 & 0 & -1/3
  \end{bmatrix}, \  b_{2 \rm c} =
  \begin{bmatrix}
    1/6 & 0 & 0 \\ \\
    0 & 1/6 & 0 \\ \\
    0 & 0 & -1/3
  \end{bmatrix}, \   b_{3 \rm c} =
  \begin{bmatrix}
    0 & 0 & 0 \\ \\
    0 & 0 & 0 \\ \\
    0 & 0 & 0
  \end{bmatrix}
\label{bcomp}
\end{equation}
An alternative to the anisotropy invariant maps is the barycentric map introduced by \citet{banerjee2007presentation,banerjee2008anisotropy}, where each realizable anisotropic state of $b_{ij}$ is written as a linear combination of the three limiting states $b_{1 \rm c}$, $b_{2 \rm c}$, and $b_{3 \rm c}$ as, \begin{equation}
C_{1 \rm c}b_{1 \rm c}+C_{2 \rm c}b_{2 \rm c}+C_{3 \rm c}b_{3 \rm c},
\label{bcomp1}
\end{equation}
where the coefficients $C_{1 \rm c}$, $C_{2 \rm c}$, and $C_{3 \rm c}$ are the three corresponding weights associated with the three limiting states, defined as:
\begin{align}
\begin{split}
C_{1 \rm c} &= e_{1}-e_{2}
\\
C_{2 \rm c} &= 2(e_{2}-e_{3})
\\
C_{3 \rm c} &= 3e_{3}+1
\end{split},
\label{bm}
\end{align}
with
\begin{equation}
C_{1 \rm c}+C_{2 \rm c}+C_{3 \rm c}=1,
\label{bm1}
\end{equation}
where $e_{1}$, $e_{2}$, $e_{3}$ are the three eigenvalues of $b_{ij}$ in the order $e_{1} > e_{2} > e_{3}$ \citep{liu2017scale,brugger2018scalewise}. Note that, these three coefficients $C_{1 \rm c}$, $C_{2 \rm c}$, and $C_{3 \rm c}$ are bounded between 0 and 1. In the extreme case, one of the coefficients taking the value 0, signifies the particular limiting state associated with that coefficient does not exist. Similarly, 1 signifies only that particular limiting state exists while the other two states being non-existent \citep{banerjee2008anisotropy}. Given the linearity in the construction of the barycentric map, it provides a non-distorted visualization of anisotropy \citep{radenkovic2014anisotropy}. \citet{banerjee2007presentation} defined the coefficient $C_{3 \rm c}$ as the degree of isotropy, such that the higher the value of $C_{3 \rm c}$ is, the anisotropic state of $b_{ij}$ is more dominated by the 3-component isotropy. The anisotropic states of $b_{ij}$ can be represented by the $RGB$ colour map of \citet{emory2014visualizing} as,
\begin{equation}
\begin{bmatrix}
  R \\
  G \\
  B \\
\end{bmatrix}
=
C_{1 \rm c}
\begin{bmatrix}
  1 \\
  0 \\
  0 \\
\end{bmatrix}
+
C_{2 \rm c}
\begin{bmatrix}
  0 \\
  1 \\
  0 \\
\end{bmatrix}
+
C_{3 \rm c}
\begin{bmatrix}
  0 \\
  0 \\
  1 \\
\end{bmatrix}
\label{bm2},
\end{equation}
such that the 1-component anisotropy is red, 2-component anisotropy is green, and the 3-component isotropy is blue. All other states within the barycentric map are linear combinations of these three colours. 

\begin{figure*}
\centering
\hspace*{-1.5in}
\includegraphics[width=1.5\textwidth]{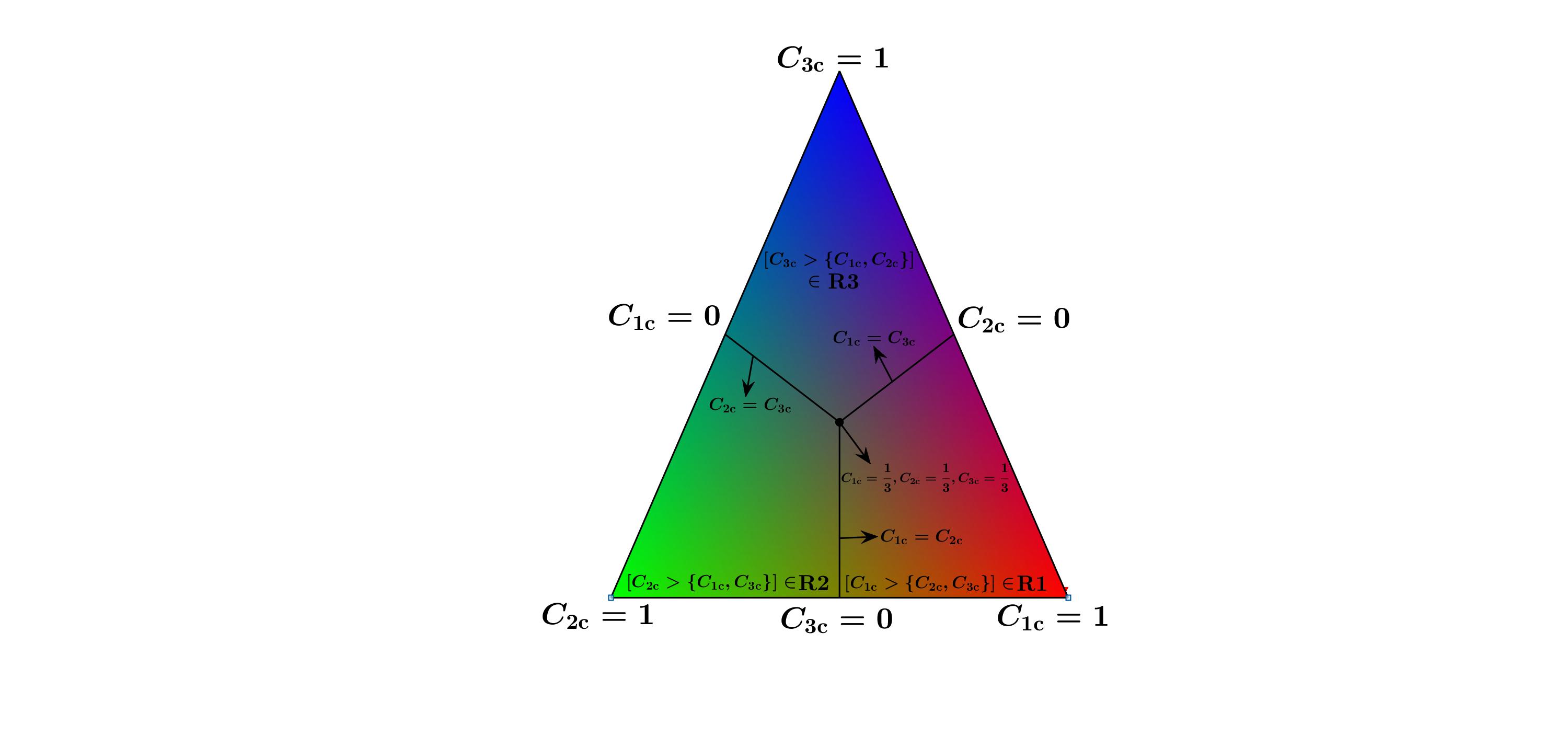}
  \caption{An example of a barycentric map spanned by an equilateral triangle is shown to graphically illustrate the anisotropic states of $b_{ij}$, using the $RGB$ colour map of \citet{emory2014visualizing}. The three vertices of the equilateral triangle represent the three limiting states with coefficients $C_{1 \rm c}$, $C_{2 \rm c}$, or $C_{3 \rm c}$ being equal to 1. At the sides opposite to the vertices any one of these coefficients are 0, which indicate the absence of that particular anisotropic state associated with it. The dark circle is the centroid of the equilateral triangle where $C_{1 \rm c}=C_{2 \rm c}=C_{3 \rm c}=$ 1/3. The three black lines are the three perpendicular bisectors which divide the equilateral triangle into three equal regions: R1 (right-third portion), R2 (left-third portion), and R3 (top-third portion). In each of these three regions (R1, R2, or R3) the anisotropic state of $b_{ij}$ is dominated by a particular limiting state associated with its coefficient $C_{1 \rm c}$, $C_{2 \rm c}$, or $C_{3 \rm c}$.}
\label{fig:1_1}
\end{figure*}

Since in this study we will be using the barycentric map to visualize the anisotropic states of $b_{ij}$, some details about its construction is appropriate here. The barycentric map is spanned by an Euclidean domain where the three limiting states of $b_{ij}$ are placed at the three vertices of an equilateral triangle having the coordinates (0, 0) for the 2-component anisotropy, (1, 0) for the 1-component anisotropy, and ($1/2$, $\sqrt{3}/2$) for the 3-component isotropy \citep{stiperski2018dependence}. For visualizing the anisotropic states, we have employed the $RGB$ colour map of \citet{emory2014visualizing} (see (\ref{bm2})). This is graphically illustrated in figure \ref{fig:1_1}. The coordinate system ($x$, $y$) of the barycentric map is defined as,
\begin{align}
\begin{split}
x &= C_{1 \rm c}x_{1 \rm c}+C_{2 \rm c}x_{2 \rm c}+C_{3 \rm c}x_{3 \rm c}
\\
&= C_{1 \rm c}+\frac{C_{3 \rm c}}{2},
\end{split}
\label{bcm_1}
\end{align}
and
\begin{align}
\begin{split}
y &= C_{1 \rm c}y_{1 \rm c}+C_{2 \rm c}y_{2 \rm c}+C_{3 \rm c}y_{3 \rm c}
\\
&= \frac{\sqrt{3}}{2}C_{3 \rm c},
\end{split}
\label{bcm_2}
\end{align}
such that the distance from the base of the equilateral triangle is directly proportional to the degree of isotropy ($C_{3 \rm c}$) of $b_{ij}$ \citep{stiperski2018dependence}. At the centroid of the barycentric map ($1/2$, $\sqrt{3}/6$), from (\ref{bcm_1}) and (\ref{bcm_2}) it can be shown that, 
\begin{equation}
C_{1 \rm c}=C_{2 \rm c}=C_{3 \rm c}=1/3.
\label{centre}
\end{equation} 
This barycentric map can also be divided into three equal regions; R1 (right-third portion), R2 (left-third portion), and R3 (top-third portion), by drawing three perpendicular bisectors from the centroid of the map (figure \ref{fig:1_1}). From (\ref{bcm_1}) and (\ref{bcm_2}), along with the constraint defined in (\ref{bm1}), these perpendicular bisectors can be represented mathematically as,
\begin{align}
\begin{split}
C_{2 \rm c}=C_{1 \rm c}
\\
C_{1 \rm c}=C_{3 \rm c}
\\
C_{2 \rm c}=C_{3 \rm c}.
\end{split}
\label{bcm_4}
\end{align}
Subsequently from symmetry it follows that, in the region R1, $C_{1 \rm c} > \{C_{2 \rm c}, C_{3 \rm c}\}$, in the region R2, $C_{2 \rm c} > \{C_{1 \rm c}, C_{3 \rm c}\}$, and in the region R3, $C_{3 \rm c} > \{C_{1 \rm c}, C_{2 \rm c}\}$. Therefore, in each of these three regions (R1, R2, or R3) the anisotropic state of $b_{ij}$ is dominated by a particular limiting state associated with its coefficient ($C_{1 \rm c}$, $C_{2 \rm c}$, or $C_{3 \rm c}$). 

\subsubsection{Representation of anisotropy on $\hat{T}$-$\hat{w}$ quadrant plane}
\label{contour_map}
To study the detailed correspondence between the anisotropic states of $b_{ij}$ and the heat flux events occurring in the $\hat{T}$-$\hat{w}$ quadrant plane, we first linearly bin $\hat{T}$ and $\hat{w}$ into a uniform 50$\times$50 grid for each run belonging to a particular stability class. The widths of each grid are defined as,
\begin{equation}
d\hat{x}=\frac{\hat{x}_{\rm max}-\hat{x}_{\rm min}}{50} \ (x=w,T)
\label{dx}.
\end{equation}

We choose the maximum ($\hat{T}_{\rm max}$, $\hat{w}_{\rm max}$) and minimum values ($\hat{T}_{\rm min}$, $\hat{w}_{\rm min}$) over all the runs from a particular stability class to ensure the same grid for individual runs. Subsequently, we find the points lying between \{$\hat{T}_{\rm bin}(m)<\hat{T}<\hat{T}_{\rm bin}(m)+d\hat{T}$, $\hat{w}_{\rm bin}(n)<\hat{w}<\hat{w}_{\rm bin}(n)+d\hat{w}$\}, where $1\leq m \leq 50$, $1\leq n \leq 50$, and $\hat{T}_{\rm bin}(m)$ and $\hat{w}_{\rm bin}(n)$ are the edges of a particular $(m,n)$ grid. For these points, we construct the anisotropy Reynolds stress tensor at $(m,n)$ grid as,
\begin{align}
\begin{split}
\langle b_{ij}\vert\{\hat{T}_{\rm bin}(m)<\hat{T}<\hat{T}_{\rm bin}(m)+d\hat{T}, \hat{w}_{\rm bin}(n)<\hat{w}<\hat{w}_{\rm bin}(n)+d\hat{w}\} \rangle &=
\\
\frac{{(\sum{{u^\prime}_{i} {u^\prime}_{j})}_{m,n}}}{{(\sum{{u^\prime}_{i} {u^\prime}_{i})}_{m,n}}} - \frac{1}{3}\delta_{ij},
\end{split}
\label{reynolds1_11}
\end{align}
conditioned on the heat flux events occurring between, 
\begin{equation*}
\{\hat{T}_{\rm bin}(m)<\hat{T}<\hat{T}_{\rm bin}(m)+d\hat{T}, \hat{w}_{\rm bin}(n)<\hat{w}<\hat{w}_{\rm bin}(n)+d\hat{w}\}
\end{equation*}
and assign it to the value \{$\hat{T}_{\rm bin}(m)$,$\hat{w}_{\rm bin}(n)$\}.

In (\ref{reynolds1_11}), the terms ${(\sum{{u^\prime}_{i} {u^\prime}_{j})}_{m,n}}$ are the contributions to the Reynolds stress tensor from each $(m,n)$ grid. The trace of $b_{ij}$ from (\ref{reynolds1_11}) can be written as, 
\begin{align}
\begin{split}
\Big[\frac{(\sum{{u^\prime}^2})_{m,n}}{(\sum{{u^\prime}^2})_{m,n}+(\sum{{v^\prime}^2})_{m,n}+(\sum{{w^\prime}^2})_{m,n}}-\frac{1}{3}\Big] &
\\
+\Big[\frac{(\sum{{v^\prime}^2})_{m,n}}{(\sum{{u^\prime}^2})_{m,n}+(\sum{{v^\prime}^2})_{m,n}+(\sum{{w^\prime}^2})_{m,n}}-\frac{1}{3}\Big] &
\\
+\Big[\frac{(\sum{{w^\prime}^2})_{m,n}}{(\sum{{u^\prime}^2})_{m,n}+(\sum{{v^\prime}^2})_{m,n}+(\sum{{w^\prime}^2})_{m,n}}-\frac{1}{3}\Big].
\end{split}
\label{reynolds1_12}
\end{align}
This sum goes to zero due to the kinetic energy term ${(\sum{{u^\prime}_{i} {u^\prime}_{i})}_{m,n}}$ appearing in the denominator. Note that, this kinetic energy is the energy contained in each $(m,n)$ grid, rather than the total kinetic energy over the whole 30-min period. This formulation is similar to the scale decomposition of $b_{ij}$, where at each scale the anisotropic Reynolds stress tensor is normalized by the kinetic energy contained in that scale to make it trace-free \citep{yeung1991response,liu2017scale,brugger2018scalewise}. 

To assess the frequency of occurrences of these heat flux events, we also compute the joint probability density function (JPDF) between $\hat{T}$ and $\hat{w}$ \citep{tennekes1972first} as,
\begin{equation}
P\Big(\hat{T}_{\rm bin}(m),\hat{w}_{\rm bin}(n)\Big)=\frac{N_{m,n}}{N \ d\hat{T} \ d\hat{w}}, 
\label{P_f}
\end{equation}
where $N_{m,n}$ is the number of points lying in $(m,n)$ grid and $N$ is the total number of points in a 30-min run (36000 for SLTEST data). Clearly, 
\begin{equation}
\int_{\hat{w}_{\rm min}}^{\hat{w}_{\rm max}} \int_{\hat{T}_{\rm min}}^{\hat{T}_{\rm max}} P\Big(\hat{T}_{\rm bin}(m),\hat{w}_{\rm bin}(n)\Big) \,d\hat{T} \,d\hat{w}=1. 
\label{P_f1}
\end{equation}
Following \citet{nakagawa1977prediction}, we also calculate the bivariate Gaussian JPDF for each grid as,
\begin{align}
\begin{split}
&G(\hat{T}_{\rm bin}(m),\hat{w}_{\rm bin}(n))=
\\
&\quad\quad \frac{1}{2 \pi \sqrt{1-{R^{2}_{wT}}}}\exp{\Big[-\Big(\frac{{\hat{T}^2_{\rm bin}(m)}-2R_{wT}\hat{T}_{\rm bin}(m)\hat{w}_{\rm bin}(n)+{\hat{w}^2_{\rm bin}(n)}}{2(1-{R^{2}_{wT}})}\Big)\Big]},
\end{split}
\label{G_f}
\end{align}
where $R_{wT}$ is the correlation coefficient between $w$ and $T$ ($\overline{w^{\prime}T^{\prime}}/\sigma_{w}\sigma_{T}$).

If the three eigenvalues of $b_{ij}$ (as defined in (\ref{reynolds1_11})) are $e_{1b}$, $e_{2b}$, and $e_{3b}$ respectively with $e_{1b}>e_{2b}>e_{3b}$, we can calculate the degree of isotropy for $(m,n)$ grid as,
\begin{equation}
\langle C_{3 \rm c}\vert\{\hat{T}_{\rm bin}(m),\hat{w}_{\rm bin}(n)\} \rangle=3e_{3b}+1,
\label{eta binned}
\end{equation}
and the $RGB$ colour map of its anisotropic states as,
\begin{align}
\begin{split}
\begin{bmatrix}
  R \\
  G \\
  B \\
\end{bmatrix}
=
\langle C_{1 \rm c}\vert\{\hat{T}_{\rm bin}(m),\hat{w}_{\rm bin}(n)\} \rangle
\begin{bmatrix}
  1 \\
  0 \\
  0 \\
\end{bmatrix}
+
\langle C_{2 \rm c}\vert\{\hat{T}_{\rm bin}(m),\hat{w}_{\rm bin}(n)\} \rangle
\begin{bmatrix}
  0 \\
  1 \\
  0 \\
\end{bmatrix} &
\\
+\langle C_{3 \rm c}\vert\{\hat{T}_{\rm bin}(m),\hat{w}_{\rm bin}(n)\} \rangle
\begin{bmatrix}
  0 \\
  0 \\
  1 \\
\end{bmatrix},
\end{split}
\label{bm3}
\end{align}
with
\begin{align}
\begin{split}
\langle C_{1 \rm c}\vert\{\hat{T}_{\rm bin}(m),\hat{w}_{\rm bin}(n)\} \rangle &= e_{1b}-e_{2b}
\\
\langle C_{2 \rm c}\vert\{\hat{T}_{\rm bin}(m),\hat{w}_{\rm bin}(n)\} \rangle &= 2(e_{2b}-e_{3b}).
\end{split}
\label{bm4}
\end{align}

Since we construct the same linear grid values of $\hat{T}$ and $\hat{w}$ for all the runs belonging to a particular stability class, we take the average of the JPDF, $C_{3 \rm c}$, and the $RGB$ colour matrices over all the individual periods. This averaging is necessary since it reduces the variability which exists from one run to another, due to the chaotic nature of turbulence. For a particular stability range, we can thus plot the averaged two-dimensional matrices of $P(\hat{T}_{\rm bin}(m),\hat{w}_{\rm bin}(n))$, $\langle C_{3 \rm c} \vert \{\hat{T}_{\rm bin}(m),\hat{w}_{\rm bin}(n)\} \rangle$, and the $RGB$ colour maps for each $(m,n)$ grid of $\hat{T}$-$\hat{w}$ quadrant plane. The contour maps of these matrices help to assess the anisotropic characteristics of the Reynolds stress tensor associated with the heat flux events of varying intensities and frequency of occurrences. While presenting the results in \S\ref{results2}, these averaged metrics are referred to as being associated with $(\hat{T},\hat{w})$, without explicitly mentioning these are the binned values. It is worth to note that the results obtained from this method are almost insensitive to the choice of the grid size. We verified this by changing the grid sizes of $\hat{T}$ and $\hat{w}$ by a factor of 2 (25$\times$25 and 100$\times$100) and repeating the calculations, with no appreciable change being noticed in the results (not shown). 

By performing the binning exercise in $\hat{w}$ and $\hat{T}$ as discussed above, we mask any time dependence and hence no information can be obtained about the time scales of the associated heat transporting events. Over the course of time these events from each quadrant can occur with a range of different time scales as they tend to exit and re-enter to their respective quadrant states \citep{kalmar2019complexity}. It is thus interesting to formulate a description of the distribution of Reynolds stress anisotropy associated with different time scales of these heat transporting events. To extract that information in addition to the quadrant analysis, we turn our attention to persistence analysis. 

\subsection{Persistence analysis}
\label{persistence}
In non-equilibrium systems, persistence is defined as the probability that the local value of a fluctuating field does not change sign up to a certain time \citep[e.g.,][]{majumdar1999persistence,bray2013persistence,ghannam2016persistence}. The concept of persistence has earlier been used by \citet{chamecki2013persistence} to study the non-Gaussian turbulence in canopy flows. He showed that an asymmetric velocity distribution inside the canopy can have very different persistent time scales for ejection and sweep events. \citet{chamecki2013persistence} also noted that the persistent time is equivalent to the inter-pulse periods between the subsequent zero crossings of the turbulent signal \citep{sreenivasan1983zero,kailasnath1993zero,bershadskii2004clusterization,cava2009effects}. We can apply this definition of persistence to the joint fluctuations in vertical velocity and temperature, to characterize the distribution of the time scales of the heat flux events from four different quadrants of $T^{\prime}$-$w^{\prime}$.  

\begin{figure*}
\centering
\vspace*{0.5in}
\hspace*{-0.8in}
\includegraphics[width=1.2\textwidth]{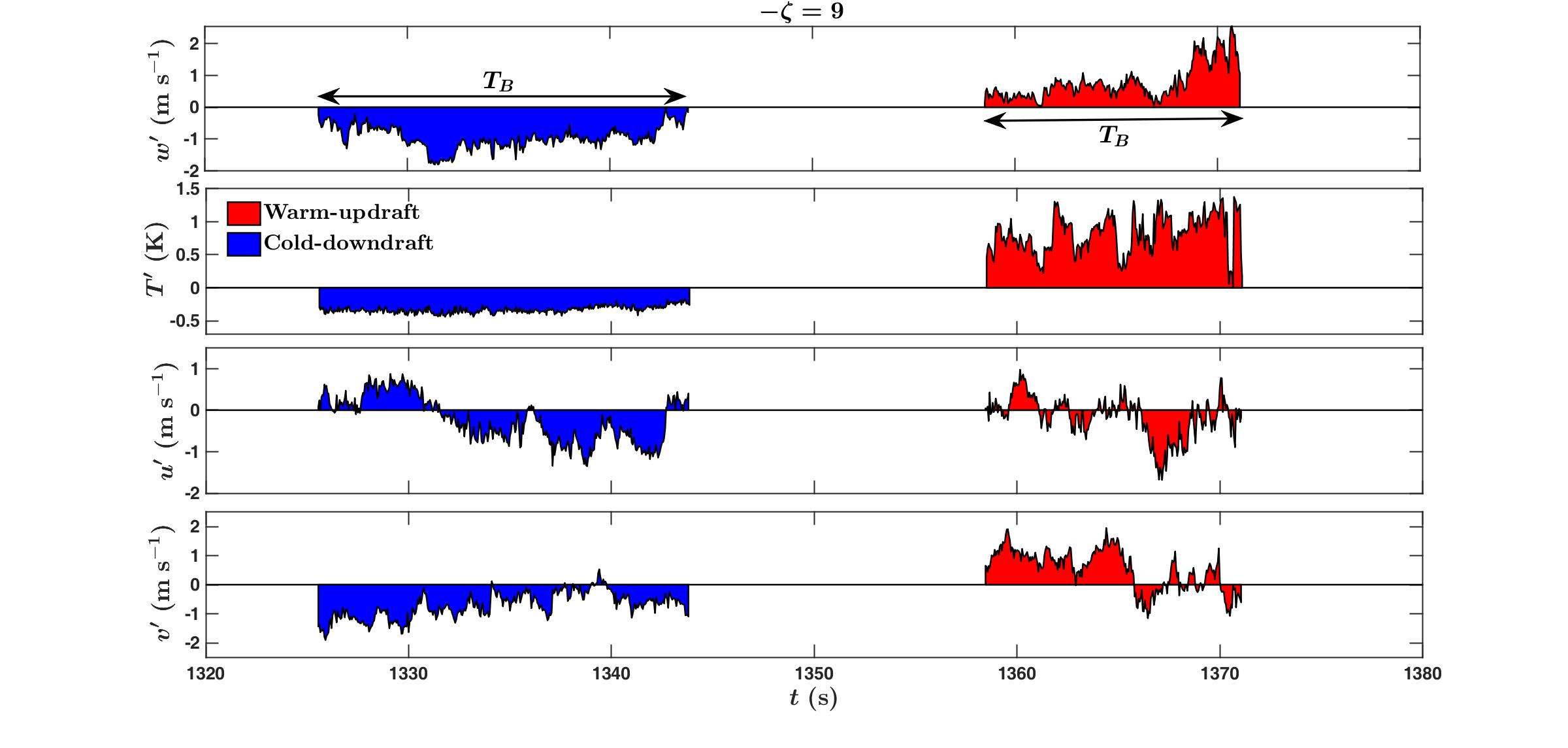}
  \caption{A 60-s long section of a time series of $w^{\prime}$, $T^{\prime}$, $u^{\prime}$, and $v^{\prime}$ for $-\zeta=$ 9. The red- and blue-shaded regions show two particular blocks of heat flux events corresponding to warm-updraft and cold-downdraft quadrants respectively, which persist for a time $T_{B}$ of around 10-20 s.}
\label{fig:1}
\end{figure*}

In order to implement our method, we choose the time series of $w^{\prime}$ and $T^{\prime}$ from any 30-min period belonging to a specific stability class (see table \ref{tab:2}), and conditionally sample the events occurring in the four different quadrants of $T^{\prime}$-$w^{\prime}$ plane. The events conditionally sampled from each quadrant of $T^{\prime}$-$w^{\prime}$ (I, II, III, or IV) can either persist as a single pulse or as a block of many consecutive pulses with a certain duration $T_{B}$, before switching to another quadrant. The duration $T_{B}$ is computed as the number of points residing within a single block, multiplied by the sampling interval of 0.05 s. In figure \ref{fig:1} we provide a graphical illustration of this method by showing a segment of a time series belonging to a particular stability range ($-\zeta=$ 9), sampled from the warm-updraft (I) and cold-downdraft (III) quadrants. The shaded blocks in figures \ref{fig:1}a to b represent warm-updrafts (red) and cold-downdrafts (blue) respectively, which persist for around 10--20 sec of duration. Associated with these blocks of warm-updrafts (red) and cold-downdrafts (blue) we also show the horizontal velocity fluctuations ($u^{\prime}$ and $v^{\prime}$) in figures \ref{fig:1}c and d.  

We convert the block duration ($T_{B}$) to a streamwise length by using the Taylor's hypothesis, that is multiplying $T_{B}$ with the mean wind speed ($\overline{u}$) computed over the 30-min period. We then scale $T_{B}\overline{u}$ with a relevant length scale. The possible candidates as the relevant length scales in an unstable ASL are the measurement height $z$ and the boundary-layer depth $z_{i}$. However, $z_{i}$ was not measured directly at SLTEST and hence an alternate large-eddy length scale $\lambda$ was used by \citet{chowdhuri2019empirical}, where $\lambda$ was computed as the peak wavelength of the horizontal velocity spectrum at $z=25.7$ m. This was based on the observation that the large-scale structures contribute directly to the horizontal velocity spectrum in the ASL \citep{kaimal1976turbulence,panofsky1977characteristics,banerjee2013logarithmic,banerjee2015revisiting}. As discussed by \citet{chowdhuri2019empirical}, a model spectrum of the form,
\begin{equation}
\kappa S_{uu}(\kappa)=\frac{a\kappa}{{(1+b\kappa)}^{\frac{5}{3}}},
\label{model}
\end{equation}
is fitted to the streamwise velocity ($u$) spectrum, where $\kappa$ is the streamwise wavenumber and $a$, $b$ are the best fit constants. By maximizing (\ref{model}) with respect to $\kappa$, $\lambda$ is evaluated as $4\pi b/3$. The other details and the rationale behind the computation of $\lambda$ can be found in \citet{chowdhuri2019empirical}.

The spectrum or the scalewise distribution of the normalized streamwise lengths of the blocks ($(T_{B}\overline{u})/\ell$, where $\ell$ can be either $z$, $\lambda$, or the combination of the two) can be at least few decades wide, given the large variation in $T_{B}$, ranging from a minimum of 0.05 s (sampling interval) to few seconds. For the blocks associated with each $T^{\prime}$-$w^{\prime}$ quadrant, we thus logarithmically bin their scaled streamwise lengths ($(T_{B}\overline{u})/\ell$) into 60 bins, where the minimum and maximum are chosen over all the 30-min periods which fall within a particular stability class. Below we discuss the method to compute the Reynolds stress anisotropy associated with these blocks of different normalized streamwise lengths. Broadly speaking, this description can be considered to be a one-dimensional analogue of the analysis carried out by \citet{dong2017coherent}. Instead of defining the sizes of the structures as connected regions in a three-dimensional space, we define those as connected points in streamwise direction after converting the temporal signals into spatial signals using Taylor's hypothesis.

\subsubsection{The distribution of the Reynolds stress anisotropy}
\label{RST_persistence}
For any particular $T^{\prime}$-$w^{\prime}$ quadrant, we collect all the blocks of the heat flux events having their normalized streamwise lengths between
\begin{equation*}
(T_{B}\overline{u}/\ell)_{\rm bin}\{m\}< (T_{B}\overline{u}/\ell)< (T_{B}\overline{u}/\ell)_{\rm bin}\{m\}+d \log{(T_{B}\overline{u}/\ell)},
\end{equation*}
where $(T_{B}\overline{u}/\ell)_{\rm bin}\{m\}$ is the logarithmically binned value, $d \log{(T_{B}\overline{u}/\ell)}$ is the bin-width, and $m$ is the index of the bin ($1\leq m \leq 60$). The bin-width is defined as, 
\begin{equation}
d\log{(T_{B}\overline{u}/\ell)}=\frac{\log{{(T_{B}\overline{u}/\ell)}_{\rm max}}-\log{{(T_{B}\overline{u}/\ell)}_{\rm min}}}{60}.
\label{binwidth}
\end{equation}

We construct the anisotropy Reynolds stress tensor associated with these blocks as,
\begin{align}
\begin{split}
\Big \langle b_{ij}\vert\Big[(T_{B}\overline{u}/\ell)_{\rm bin}\{m\}< (T_{B}\overline{u}/\ell)< (T_{B}\overline{u}/\ell)_{\rm bin}\{m\}+d \log{(T_{B}\overline{u}/\ell)}\Big] \Big \rangle &=
\\
\frac{{\sum{{u^\prime}_{i} {u^\prime}_{j}}}}{\sum{{u^\prime}_{i} {u^\prime}_{i}}} - \frac{1}{3}\delta_{ij},
\end{split}
\label{reynolds2_11}
\end{align}
and assign it to a streamwise size of $(T_{B}\overline{u}/\ell)_{\rm bin}\{m\}$. In (\ref{reynolds2_11}), the terms $\sum{{u^\prime}_{i} {u^\prime}_{j}}$ are the contributions to the Reynolds stress tensor from all the blocks having their sizes between $(T_{B}\overline{u}/\ell)_{\rm bin}\{m\}$ and $(T_{B}\overline{u}/\ell)_{\rm bin}\{m\}+d \log{(T_{B}\overline{u}/\ell)}$. 

Similar to \S\ref{contour_map}, we calculate the three coefficients associated with 1-component anisotropy, 2-component anisotropy, and 3-component isotropy ($C_{1 \rm c}$, $C_{2 \rm c}$, and $C_{3 \rm c}$) of $\langle b_{ij}\vert(T_{B}\overline{u}/\ell)_{\rm bin}\{m\} \rangle$ as,
\begin{align}
\begin{split}
\langle C_{1 \rm c}\vert(T_{B}\overline{u}/\ell)_{\rm bin}\{m\} \rangle &= {\tilde{e}}_{1b}-{\tilde{e}}_{2b}
\\
\langle C_{2 \rm c}\vert(T_{B}\overline{u}/\ell)_{\rm bin}\{m\} \rangle &= 2({\tilde{e}}_{2b}-{\tilde{e}}_{3b})
\\
\langle C_{3 \rm c}\vert(T_{B}\overline{u}/\ell)_{\rm bin}\{m\} \rangle &=
3{\tilde{e}}_{3b}+1,
\end{split}
\label{bm4_1}
\end{align}
where ${\tilde{e}}_{1b}$, ${\tilde{e}}_{2b}$, and ${\tilde{e}}_{3b}$ are the three eigenvalues of $\langle b_{ij}\vert(T_{B}\overline{u}/\ell)_{\rm bin}\{m\} \rangle$ with ${\tilde{e}}_{1b}>{\tilde{e}}_{2b}>{\tilde{e}}_{3b}$. 

Since we construct the same logarithmic grids of $(T_{B}\overline{u})/\ell$ for all the runs belonging to a particular stability class, we take the average of these three coefficients over all these periods to reduce the run-to-run variability. 

\subsubsection{Probability and flux distributions}
\label{Prob_flux_persistence}
The probability density function (PDF) of the normalized streamwise lengths of the blocks belonging to any particular $T^{\prime}$-$w^{\prime}$ quadrant is calculated as, 
\begin{equation}
P(T_{B}\overline{u}/\ell)_{\rm bin}\{m\}=\frac{N_{\rm b}}{N_{\rm tot} \ d \log{(T_{B}\overline{u}/\ell)}},
\label{P_f_1}
\end{equation}
where $N_{\rm b}$ is the number of blocks lying between,
\begin{equation*}
(T_{B}\overline{u}/\ell)_{\rm bin}\{m\}< (T_{B}\overline{u}/\ell)< (T_{B}\overline{u}/\ell)_{\rm bin}\{m\}+d \log{(T_{B}\overline{u}/\ell)},
\end{equation*}
and $N_{\rm tot}$ is the total number of blocks detected over a 30-min period (from the same quadrant). Clearly,
\begin{equation}
\int_{(T_{B}\overline{u}/\ell)_{\rm min}}^{{(T_{B}\overline{u}/\ell)_{\rm max}}} \Big[P(T_{B}\overline{u}/\ell)_{\rm bin}\Big] \, d \log{(T_{B}\overline{u}/\ell)}=1.
\label{P_f_2}
\end{equation} 

The heat and momentum fluxes within these blocks are defined as,
\begin{align}
\begin{split}
\Big \langle w^{\prime}x^{\prime}\vert\Big[(T_{B}\overline{u}/\ell)_{\rm bin}\{m\}< (T_{B}\overline{u}/\ell)< (T_{B}\overline{u}/\ell)_{\rm bin}\{m\}+d \log{(T_{B}\overline{u}/\ell)}\Big] \Big \rangle &=
\\
\frac{\sum{w^{\prime}x^{\prime}}}{N \times d \log{(T_{B}\overline{u}/\ell)}}, \ (x=u,T)
\end{split}
\label{F_1}
\end{align}
where $N$ is the number of samples in a 30-min run for the SLTEST data. These heat and momentum flux distributions are scaled by the product of the standard deviations $\sigma_{w}\sigma_{T}$ and $\sigma_{u}\sigma_{w}$, respectively. When these scaled flux distributions from (\ref{F_1}) are integrated over the whole spectrum of $(T_{B}\overline{u})/\ell$, the results show the strength of the coupling between $w^{\prime}$ and $T^{\prime}$ ($u^{\prime}$) from each quadrant X of $T^{\prime}$-$w^{\prime}$ (X $=$ I, II, III, or IV) as,
\begin{equation}
\int_{(T_{B}\overline{u}/\ell)_{\rm min}}^{{(T_{B}\overline{u}/\ell)_{\rm max}}} \Big[\frac{{\langle w^{\prime}x^{\prime}\vert(T_{B}\overline{u}/\ell)}_{\rm bin}\rangle}{\sigma_{w}\sigma_{x}}\Big] \, d \log{(T_{B}\overline{u}/\ell)}= \Big(\frac{\overline{w^{\prime}x^{\prime}}}{\sigma_{w}\sigma_{x}}\Big)_{\rm X}, \ (x=u,T).
\label{F_2}
\end{equation} 

Similar to \S\ref{RST_persistence}, we take the average of the PDFs and the heat and momentum flux distributions over all the 30-min periods belonging to a particular stability class. While presenting the results in \S\ref{results3}, these averaged distributions of the degree of isotropy, probability, and fluxes are referred to being associated with $(T_{B}\overline{u})/\ell$ only, without explicitly mentioning these are the binned values. Apart from that, the amount of spread between the individual 30-min runs for a particular stability class is computed as one standard deviation from the ensemble average and shown as error-bars. Similar to quadrant analysis, results obtained from this method have also been verified for sensitivity to the choice of the number of bins.

\section{Results and discussion}\label{results}
We begin with discussing the general characteristics of the Reynolds stress anisotropy with the change in the stability ratio $-\zeta$. We also highlight the correspondence between the intermittent nature of turbulent heat transport and the Reynolds stress anisotropy. By presenting the relevant results, this correspondence is further investigated in details, complemented with the quadrant and persistence analyses of the heat flux events. The possible physical interpretations of these results are also discussed. 

\begin{figure*}
\centering
\vspace*{0.5in}
\hspace*{-0.8in}
\includegraphics[width=1.2\textwidth]{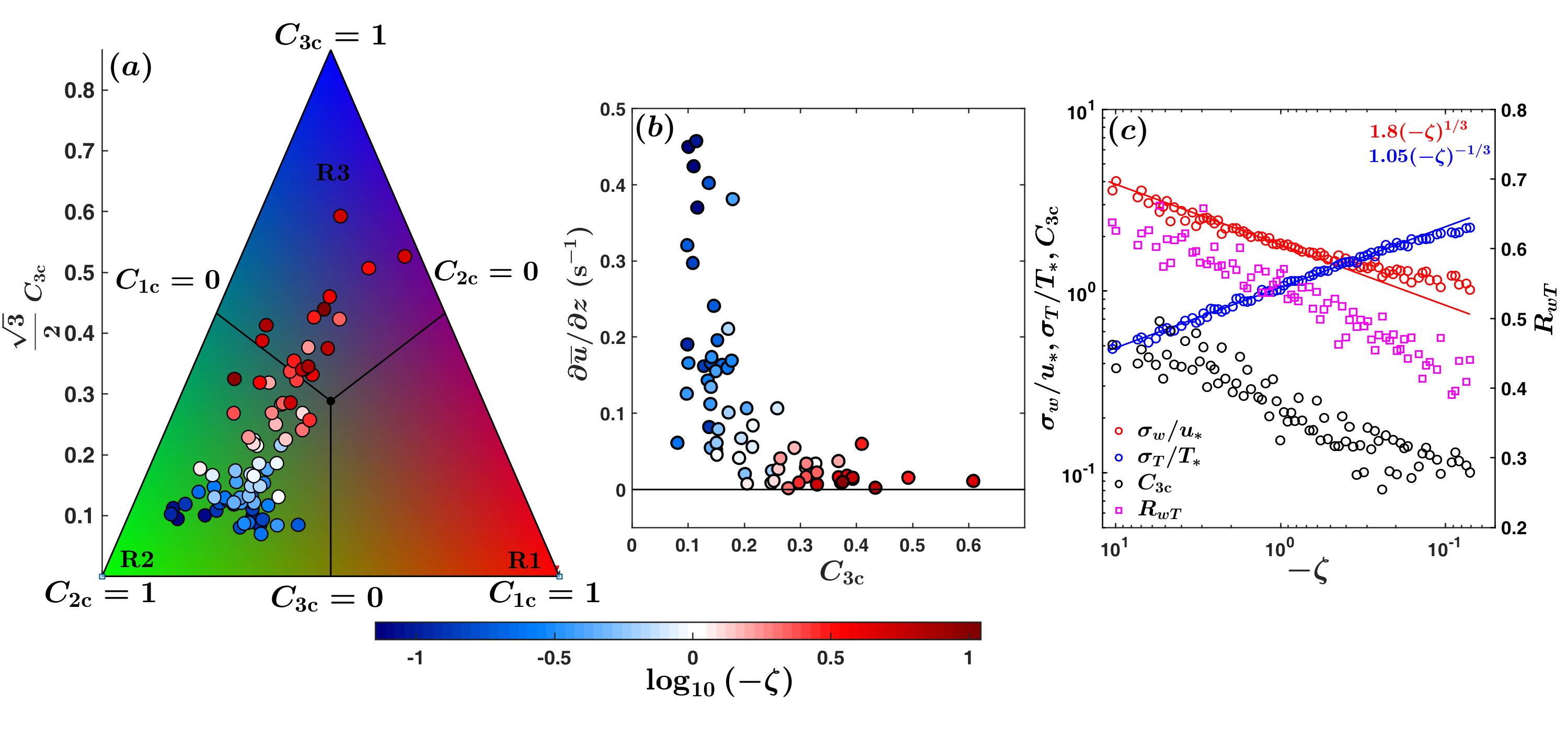}
  \caption{The plots of the (a) anisotropic states of $b_{ij}$ on the barycentric map (see (\ref{bcm_1}) and (\ref{bcm_2})), (b) the degree of isotropy $C_{3 \rm c}$ (see (\ref{bm})) with the wind shear ($\partial{\overline{u}}/\partial z$), and (c) scaled vertical velocity and temperature standard deviations ($\sigma_{w}/u_{*}$ and $\sigma_{T}/T_{*}$), $C_{3 \rm c}$, correlation coefficient between the vertical velocity and temperature ($R_{wT}$) are shown against the stability ratio ($-\zeta$). In panel (c), the left $y$ axis is logarithmic, the right $y$ axis is linear, and the $x$ axis is reversed such that the $-\zeta$ values proceed from large to small. The thick blue and red lines denote the local free convection scalings for $\sigma_{w}/u_{*}$ and $\sigma_{T}/T_{*}$. The colour bar at the bottom corresponds to both the panels (a, b), showing the stability ratios as $\log_{10}(-\zeta)$.}
\label{fig:2}
\end{figure*}

\subsection{The characteristics of Reynolds stress anisotropy with stability}\label{results1}
We discuss the general effect of stability on the Reynolds stress anisotropy associated with the 30-min averaged flow in an unstable ASL. We establish that along with stability, the Reynolds stress anisotropy in the averaged flow is also related to the intermittent and asymmetric nature of turbulent heat transport.

Figure \ref{fig:2}a shows the anisotropic states of the Reynolds stress tensor for the 30-min averaged flow plotted on the barycentric map (see (\ref{bm2}), (\ref{bcm_1}), and (\ref{bcm_2})) with the stability ratio $-\zeta$. The variation of the three associated coefficients ($C_{1 \rm c}$, $C_{2 \rm c}$, and $C_{3 \rm c}$) with $-\zeta$ are provided in figure S1a (supplementary material). As shown in figure \ref{fig:1_1}, the barycentric map is spanned by an equilateral triangle which can be divided into three regions R1, R2, and R3 where the anisotropic states of $b_{ij}$ are dominated by 1-component anisotropy, 2-component anisotropy, and 3-component isotropy respectively. As evident from figure \ref{fig:2}a, the anisotropic states of $b_{ij}$ move towards the region R3 from the region R2 as $-\zeta$ approaches the local free convection limit ($-\zeta>1$). This implies that the anisotropic state of $b_{ij}$ is more dominated by the 3-component isotropy as the surface layer becomes highly-convective. The reason for this is, in a highly-convective surface layer the turbulent kinetic energy (TKE) is mainly generated in the vertical direction through buoyancy production term, while in the horizontal direction the production of TKE due to shear is almost negligible. However, the pressure-strain correlation in highly-convective conditions efficiently redistribute the TKE generated in vertical to the horizontal direction, thus driving the turbulence to be dominated by the 3-component isotropy \citep{mcbean1975vertical,zhuang1995dynamics,bou2018role}. The reason behind the effectiveness of pressure-strain correlation in redistributing the TKE in highly-convective conditions is related to the covariance between the pressure and vertical velocity fluctuations, as detailed in the physical model of \citet{mcbean1975vertical}. On the other hand, for small values of $-\zeta$ ($0<-\zeta<0.2$) the anisotropic state of $b_{ij}$ is dominated by the 2-component anisotropy, as the blue shaded points in figure \ref{fig:2}a remain concentrated within the region R2. From table \ref{tab:2}, it is clear that the near-neutral stability class ($0<-\zeta<0.2$) corresponds to the lowest three levels of the SLTEST experiment ($z=$ 1.4, 2.1, 3 m), where due to the blocking of the ground the vertical velocity fluctuations are suppressed. Therefore, the turbulence very close to the ground is in a 2-component anisotropic state dominated by the horizontal velocity components. This is in agreement with the studies by \citet{krogstad2000invariant} and \citet{ali2018anisotropy}. 

Apart from the anisotropic states of $b_{ij}$, we can also evaluate its degree of isotropy $C_{3 \rm c}$ to quantify how closer the turbulence is towards the 3-component isotropy \citep{banerjee2007presentation,banerjee2008anisotropy}. From figure \ref{fig:2}b we note that for $0<-\zeta<0.2$, strong anisotropic turbulence ($C_{3 \rm c} \approx 0.1$) is associated with large wind shear ($\partial{\overline{u}}/\partial{z}$). The wind shear is approximated using the finite-difference scheme \citep{arya2001introduction,stull2012introduction} as, 
\begin{equation}
\Big(\frac{\partial{\overline{u}}}{\partial{z}}\Big)_{z_{\rm m}} \approx \frac{\overline{u}(z_{2})-\overline{u}(z_{1})}{z_{2}-z_{1}},
\label{windshear}
\end{equation}
where $z_{2}$ and $z_{1}$ are the two adjacent levels from the SLTEST data with $z_{2}>z_{1}$ and $z_{\rm m}=(z_{2}+z_{1})/2$. However, in the limit of local free convection ($-\zeta>$ 1) the effect of wind shear is weak and the turbulence is more dominated by 3-component isotropy ($C_{3 \rm c}>1/3$). This result is consistent with the observations of \citet{stiperski2018dependence} where they found that in an unstable surface layer as we approach $z \to 0$ (associated with small values of $-\zeta$) the anisotropic characteristics of turbulence are dominated by strong wind shear. However as the local free convection is approached ($-\zeta>$ 1), the effect of wind shear weakens and the turbulence becomes less anisotropic. It is also in agreement with \citet{jin2003equilibrium}, where they showed both analytically and from numerical simulations that in a buoyant shear flow, the effect of increase (decrease) in buoyancy (shear) was to drive the turbulence towards isotropy. 

To investigate this further, figure \ref{fig:2}c shows the scatter plot of the scaled vertical and temperature standard deviations ($\sigma_{w}/u_{*}$ and $\sigma_{T}/T_{*}$) along with the correlation coefficient between $w$ and $T$ ($R_{wT}$) and the degree of isotropy ($C_{3 \rm c}$), against the stability ratio $-\zeta$. Note that, the temperature scale ($T_{*}$) is defined here as $H_{0}/u_{*}$ with the omission of the negative sign, to keep the quantity $\sigma_{T}/T_{*}$ positive. The local free convection scalings for $\sigma_{w}/u_{*}$ and $\sigma_{T}/T_{*}$ are given as,
\begin{align}
\begin{split}
\frac{\sigma_{w}}{u_{*}} &= 1.8 {(-\zeta)}^{1/3}
\\
\frac{\sigma_{T}}{T_{*}} &= 1.05 {(-\zeta)}^{-1/3},
\end{split}
\label{lfc}
\end{align}
where the coefficients are fitted from the data and match well with the values reported by \citet{wyngaard1971local} and \citet{monin2013statistical}. It is interesting to note that after $-\zeta<$ 0.5, the local free convection scaling does not hold for $\sigma_{w}/u_{*}$, but it extends for $\sigma_{T}/T_{*}$. \citet{khanna1997analysis} explained this as, the buoyancy-induced motions contribute more to the temperature fluctuations than the shear-induced motions. 

In an unstable surface layer, the horizontal velocity variances depend on the global stability ratio $-z_{i}/L$, rather than on $-\zeta$ \citep{monin2013statistical,panofsky1974atmospheric,panofsky1977characteristics,wyngaard2010turbulence}. Therefore, the variation in degree of isotropy ($C_{3 \rm c}$) with $-\zeta$ is mainly determined by the strength of the vertical velocity fluctuations ($\sigma_{w}$), decreasing from $C_{3 \rm c} \approx 0.6$ to $C_{3 \rm c} \approx 0.1$ as $\sigma_{w}/u_{*}$ decreases with $-\zeta$ (figure \ref{fig:2}c). From figure \ref{fig:2}c and figure S1 (supplementary material), we note that $w^{\prime}$ is more strongly coupled to $T^{\prime}$ than to $u^{\prime}$ in the local free convection ($R_{wT}\approx$ 0.65 and $R_{uw}\approx -0.05$). However, with decrease in $-\zeta$, the correlation coefficient between $w^{\prime}$ and $u^{\prime}$ increases ($R_{uw}\approx -0.25$) whereas it decreases between $w^{\prime}$ and $T^{\prime}$ ($R_{wT}\approx$ 0.4). This is also reflected in the transport efficiencies of heat ($\eta_{wT}$) and momentum ($\eta_{uw}$) defined as,
\begin{equation}
\eta_{wx}=\frac{({\sum{w^{\prime}x^{\prime}}})_{\rm \var{down-gradient}}+({\sum{w^{\prime}x^{\prime}}})_{\rm \var{counter-gradient}}}{({\sum{w^{\prime}x^{\prime}}})_{\rm \var{down-gradient}}},
\label{trnseff}
\end{equation}
where $x$ can be either $u$ or $T$ \citep{li2011coherent,bou2018role}. From figure S1b (supplementary material), it is evident that in local free convection $\eta_{uw}\to 0$ whereas $\eta_{wT}$ almost approaches a constant value of 0.9. However, with decrease in $-\zeta$, $\eta_{uw}$ increases to $\approx$ 0.6 and $\eta_{wT}$ decreases to $\approx$ 0.75. We next investigate the PDFs of $T^{\prime}$, $w^{\prime}$, and $u^{\prime}$ to establish a correspondence between the turbulence anisotropy and its transport characteristics.  

\begin{figure*}
\centering
\vspace*{0.5in}
\hspace*{-0.8in}
\includegraphics[width=1.2\textwidth]{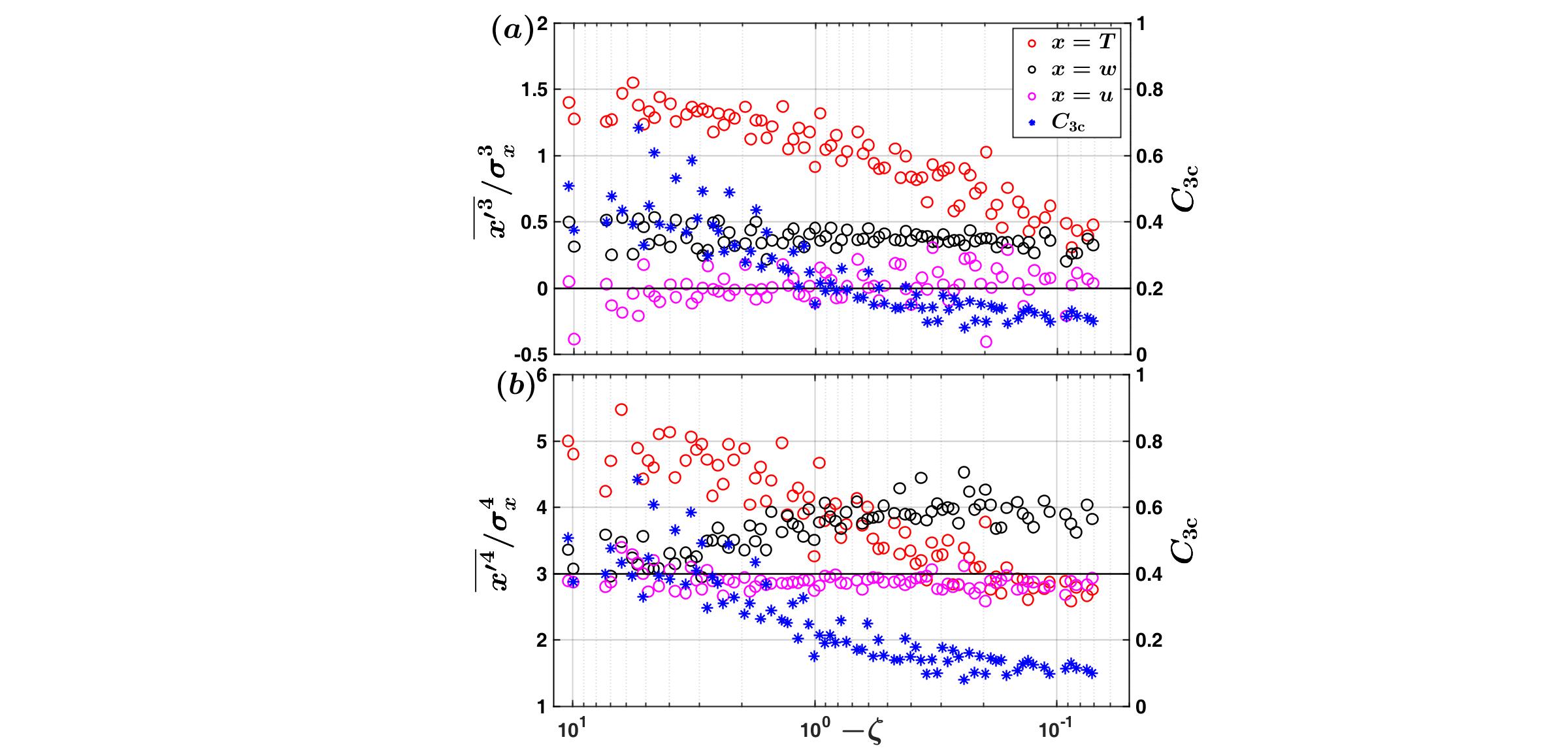}
  \caption{The scatter plot of the (a) skewness and (b) kurtosis of the temperature, vertical velocity, and streamwise velocity fluctuations ($T^{\prime}$, $w^{\prime}$, and $u^{\prime}$) are shown against $-\zeta$. The red, blue, and pink coloured open circles denote $T^{\prime}$, $w^{\prime}$ and $u^{\prime}$ respectively, with their skewness and kurtosis being plotted on the left hand side of the $y$ axis. The black stars show the degree of isotropy ($C_{3 \rm c}$, see (\ref{bm})) with its values being plotted on the right hand side of the $y$ axis. The thick horizontal black lines denote the values of 0 and 3, which are the skewness and kurtosis for the Gaussian distribution.}
\label{fig:3}
\end{figure*}

Figures \ref{fig:3}a and b show the skewness and kurtosis of the PDFs of $T^{\prime}$, $w^{\prime}$, and $u^{\prime}$ ($\overline{{x^{\prime}}^3}/\sigma_{x}^3$ and $\overline{{x^{\prime}}^4}/\sigma_{x}^4$, where $x=u,w,T$) along with the degree of isotropy ($C_{3 \rm c}$). The associated PDFs are shown in figure S2 (supplementary material). For a perfect Gaussian distribution, the skewness and kurtosis have values of 0 and 3 respectively \citep[e.g.,][]{lumley2007stochastic}. Physically, the skewness is associated with the asymmetry in the PDFs whereas the kurtosis is related to intermittency \citep{tennekes1972first,davidson2015turbulence,pouransari2015statistical}.

From figures \ref{fig:3}a and b, it is clear that the skewness and kurtosis of the temperature fluctuations are strongly non-Gaussian ($\approx$ 1.5 and 5 respectively) in the local free-convection limit ($-\zeta>$ 1). The strong non-Gaussian nature of temperature fluctuations in highly-unstable condition is remarkably consistent with the previous studies in the ASL \citep{chu1996probability,liu2011probability,garai2013interaction,lyu2018high}. Similar behaviour has also been observed in turbulent Rayleigh B\'{e}nard convection experiments of \citet{adrian1986turbulent}, \citet{balachandar1991probability} and \citet{wang2019turbulent}. The strong non-Gaussianity in $T^{\prime}$ in highly-unstable conditions is caused due to the intermittent bursts associated with warm-updrafts, interspersed with relatively more frequent quiescent cold-downdrafts bringing well-mixed air from aloft \citep{adrian1986turbulent,chu1996probability}. However, the skewness and kurtosis of $T^{\prime}$ become closer to Gaussian (0.5 and 3 respectively) for the near-neutral stability (0 $<-\zeta<$ 0.2). The close-to-Gaussian characteristics of the $T^{\prime}$ PDFs in a near-neutral ASL are in agreement with \citet{chu1996probability} and with the pipe flow experiment of \citet{nagano1988statistical} where temperature behaved more like a passive scalar.  

On the other hand, the PDFs of $u^{\prime}$ remain near-Gaussian for all the values of $-\zeta$ with its skewness and kurtosis approaching 0 and 3 respectively. For $w^{\prime}$, the skewness stays almost constant at 0.4 to 0.5 for all the values of $-\zeta$, implying the consistent upward transport of vertical kinetic energy \citep{chiba1978stability,hunt1988eddy}. However, the kurtosis for $w^{\prime}$ increases from 3 to 4 as $-\zeta$ decreases. This observation is consistent with \citet{chu1996probability} where they found the kurtosis in $w^{\prime}$ increased from 3.12 in highly-unstable conditions to 3.77 in near-neutral conditions. \citet{chiba1984note} postulated that this increase in the kurtosis of $w^{\prime}$ at small $-\zeta$ values is related to the increasing importance of the small-scale eddies near the ground. However, \citet{hong2004turbulence} hypothesized it to be related to the low-speed streaks, initiating inactive and active turbulence interactions with increasing intermittency. 

We note that the degree of isotropy ($C_{3 \rm c}$) also decreases in a similar way as the skewness and kurtosis of the temperature fluctuations approach a near-Gaussian distribution with decrease in $-\zeta$ (figure \ref{fig:3}). \citet{katul1997turbulent_b} demonstrated that the temperature skewness was directly related to the difference in the time fractions ($\Delta T_{f}$) of the warm-updraft and cold-downdraft events (asymmetry) as,
\begin{equation}
\Delta T_{f}=\frac{Q_{3}}{3\sqrt{2\pi}},
\label{skew}
\end{equation}
where $Q_{3}=\overline{{T^{\prime}}^3}/\sigma_{T}^3$, by assuming that the time fractions spent in the counter-gradient quadrants of $T^{\prime}$-$w^{\prime}$ plane could be ignored. This implies that the asymmetry in the distributions of the warm-updraft and cold-downdraft events associated with the skewness of the temperature fluctuations, has a strong correspondence with the anisotropy in the Reynolds stress tensor. In figure S3 (supplementary material) we show the heat flux fractions ($F_{f}$) and the time fractions ($T_{f}$) associated with each quadrant of $T^{\prime}$-$w^{\prime}$ plane. It indicates that in highly-unstable conditions ($-\zeta>$ 1) the warm-updrafts carry more heat flux even though they spend less time than the cold-downdrafts (see figure S3c in supplementary material). 

\begin{figure*}
\centering
\vspace*{0.5in}
\hspace*{-0.5in}
\includegraphics[width=1.2\textwidth]{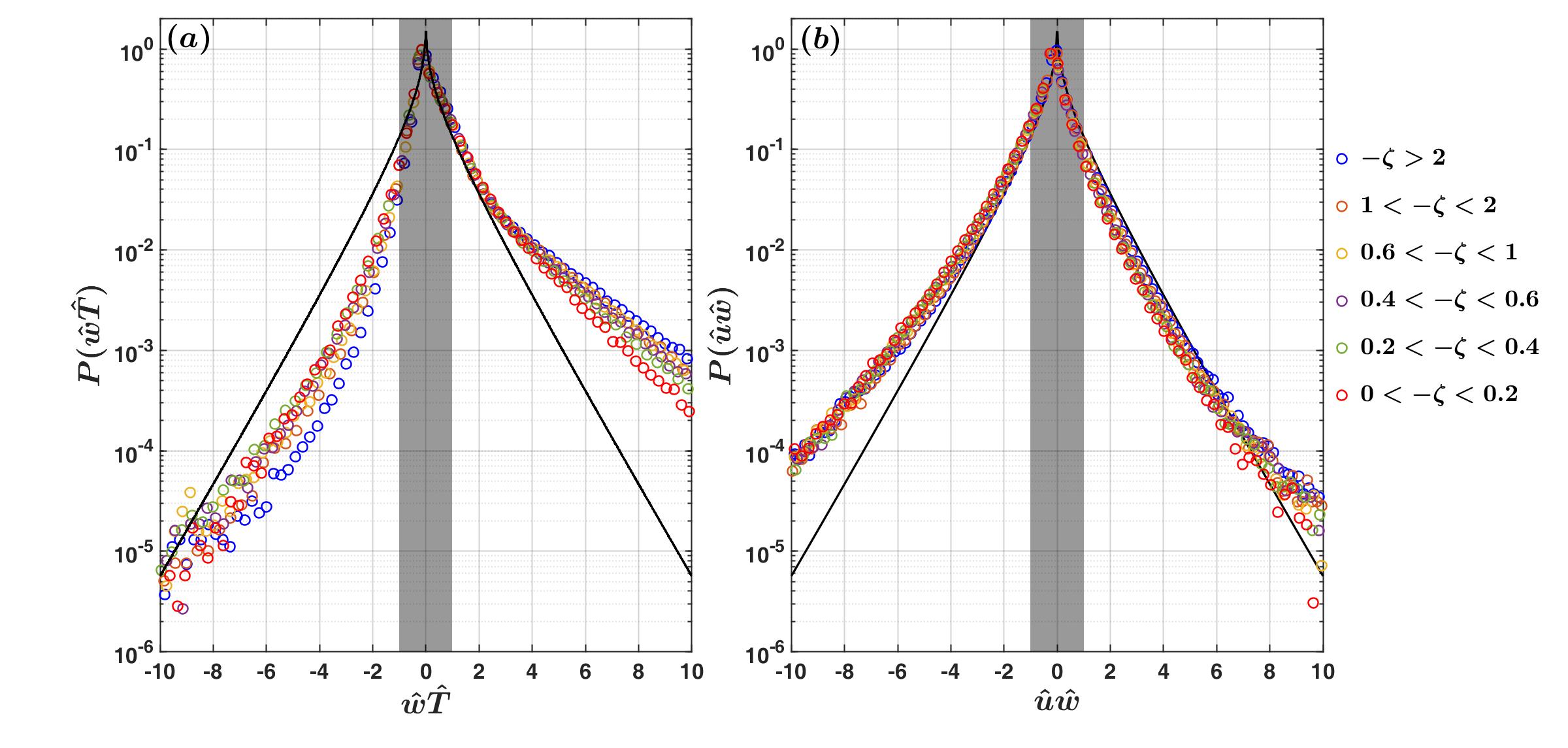}
  \caption{The PDFs of the scaled (a) heat flux ($P(\hat{w}\hat{T})$) and (b) momentum flux ($P(\hat{u}\hat{w})$), are shown for the six different classes of $-\zeta$ as indicated in the legend at the right most corner. The thick black curves denote the modified Bessel function of the second kind, which corresponds to the PDF of $\hat{w}\hat{x}$ ($x=u,T$), if $\hat{u}$, $\hat{w}$, and $\hat{T}$ were all standard Gaussian random variables (see (\ref{mbf})). The grey shaded portions show the hyperbolic hole, defined as $|\hat{x}\hat{w}|=1$ ($x=u,T$).}
\label{fig:4}
\end{figure*}

The same observation can also be made from figure \ref{fig:4}a, where the strong non-Gaussianity in the temperature fluctuations in highly-unstable conditions ($-\zeta>2$) introduces a large asymmetry in the PDFs of the scaled heat flux ($P(\hat{w}\hat{T})$). The intermittent bursts associated with warm-updrafts characterized by large kurtosis carry more heat flux than predicted by the distribution if $\hat{w}$ and $\hat{T}$ were both standard Gaussian random variables. According to \citet{krogstad2013development}, the PDF of the product of two standard Gaussian random variables $\hat{x}$ and $\hat{y}$ can be expressed as, 
\begin{equation}
P(\hat{x}\hat{y})=\frac{K_{0}\Big(|\hat{x}\hat{y}|\Big)}{\pi},
\label{mbf}
\end{equation}
where $K_{0}(|\hat{x}\hat{y}|)$ is the modified Bessel function of the second kind. However, this strong non-Gaussianity is not felt in the PDFs of the scaled momentum flux ($P(\hat{u}\hat{w})$) as the probability distributions of $u$ and $w$ fluctuations are closer to Gaussian compared to temperature (figure \ref{fig:4}b and figure S2).

In a nutshell from figures \ref{fig:3} and \ref{fig:4} one can infer that the characteristics of the turbulent heat transport in an unstable surface layer is strongly (weakly) non-Gaussian for highly (feebly) convective conditions, associated with less (more) anisotropic turbulence. However, till now we have presented the anisotropic characteristics of the averaged flow which comprises of the heat flux events from all the four quadrants of $T^{\prime}$-$w^{\prime}$. Given the asymmetric and intermittent nature of turbulent heat transport, it is thus imperative to employ event based analysis to investigate ``\emph{whether the strongest (weakest) heat flux events are associated with less (more) anisotropic turbulence}?''. Therefore, we turn our attention towards the quadrant analysis to deduce the anisotropic characteristics of the Reynolds stress tensor, associated with the heat flux events from the four different quadrants. 

\subsection{Quadrant analysis of Reynolds stress anisotropy}\label{results2}
\begin{figure*}
\centering
\vspace*{0.5in}
\hspace*{-1.2in}
\includegraphics[width=1.35\textwidth]{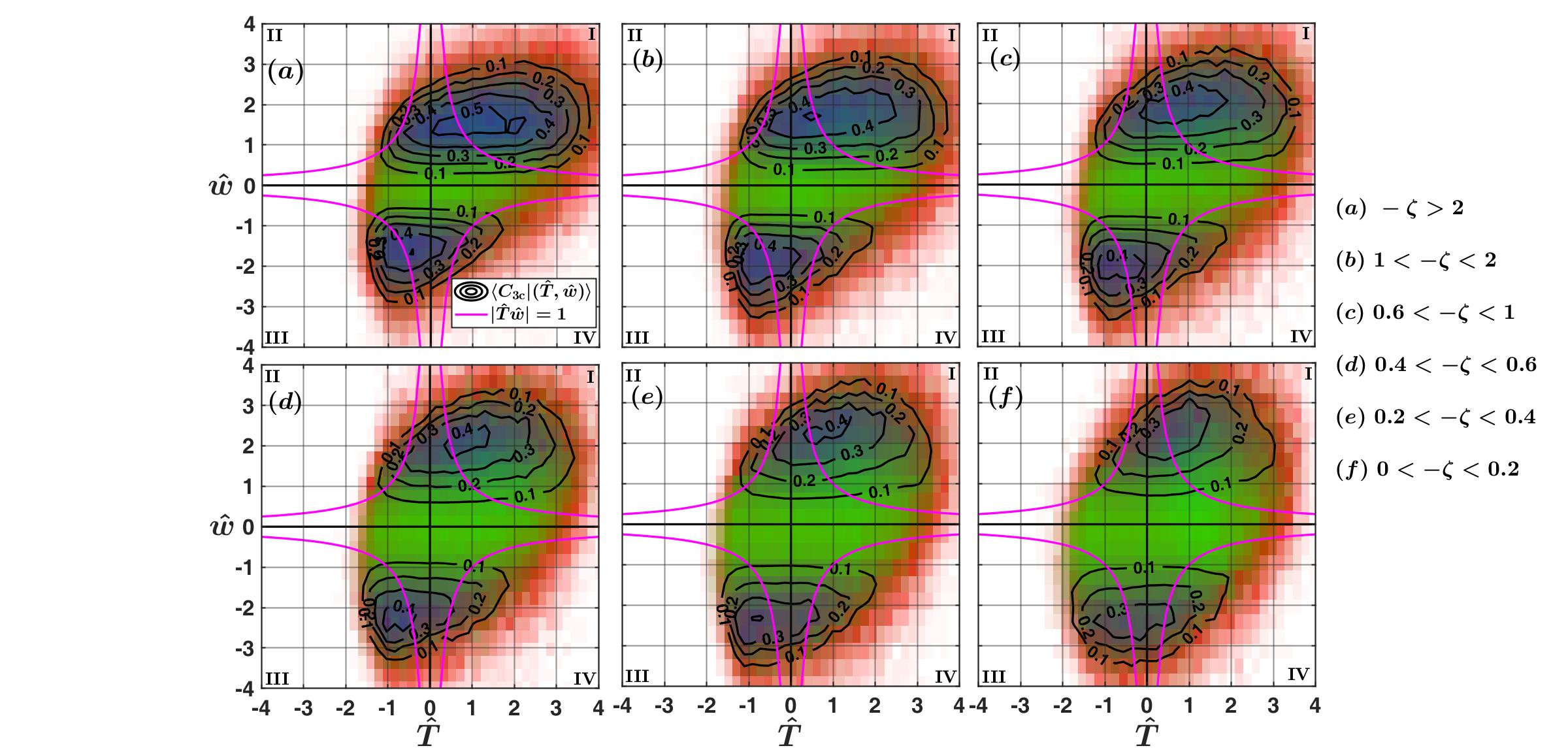}
  \caption{The quadrant maps of the degree of isotropy ($\langle C_{3 \rm c}\vert(\hat{T},\hat{w}) \rangle$, see (\ref{eta binned})) plotted on the $\hat{T}$-$\hat{w}$ quadrant plane, are shown for the six different classes of the stability ratios as indicated in the legend at the right-most corner. The anisotropic states of $\langle b_{ij}\vert(\hat{T},\hat{w}) \rangle$ are represented by the $RGB$ colour map such that the red, green, and blue shaded regions of the quadrant plane are dominated by 1-component anisotropy, 2-component anisotropy, and 3-component isotropy respectively (see (\ref{bm3})). The thick pink lines denote the hyperbolic hole $|\hat{T}\hat{w}|=1$. The quadrants I and III represent the warm-updrafts and cold-downdrafts, whereas II and IV represent the cold-updrafts and warm-downdrafts.}
\label{fig:5}
\end{figure*}

From quadrant analysis, we study the detailed correspondence between the Reynolds stress anisotropy and the heat flux events of varying intensities with their frequency of occurrences. Figure \ref{fig:5} shows the $RGB$ colour map computed by (\ref{bm3}) with the superposed contours of degree of isotropy (see (\ref{eta binned})) on $\hat{T}$-$\hat{w}$ quadrant plane. From the $RGB$ colour map, the anisotropic states of the Reynolds stress tensor in the red, green, and blue shaded regions of the $\hat{T}$-$\hat{w}$ quadrant plane are dominated by 1-component anisotropy, 2-component anisotropy, and 3-component isotropy respectively. We also include the hyperbolic hole, defined as $|\hat{T}\hat{w}|=1$, to identify the strong heat flux producing events which lie in the region outside of it \citep{smedman2007heat}. The six different panels in figures \ref{fig:5}a--f correspond to the six different stability classes as mentioned in table \ref{tab:2}. 

\begin{figure*}
\centering
\vspace*{0.5in}
\hspace*{-1.2in}
\includegraphics[width=1.35\textwidth]{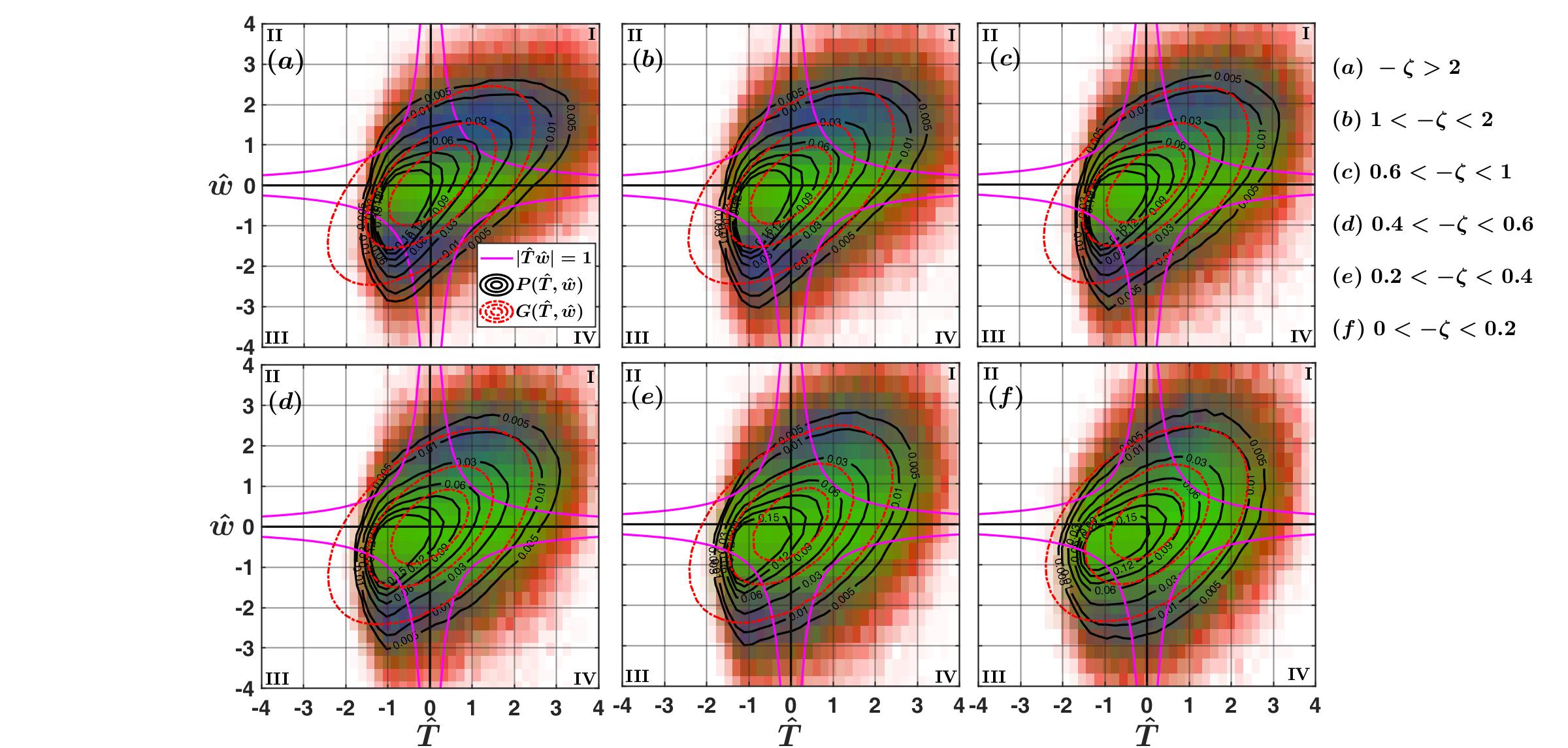}
  \caption{The contour maps of the JPDFs between $\hat{T}$ and $\hat{w}$ ($P(\hat{T},\hat{w})$ as thick black lines, see (\ref{P_f})) and the bivariate Gaussian distribution ($G(\hat{T},\hat{w})$ as dotted red lines, see (\ref{G_f})) are shown for the six different classes of stability ratios as indicated in the legend at the right most corner. The same $RGB$ colour map and hyperbolic hole from figure \ref{fig:5} are shown here too.}
\label{fig:6}
\end{figure*}

From figure \ref{fig:5}a, we notice that in highly-convective conditions ($-\zeta>2$), the anisotropic states of the Reynolds stress tensor for strong heat flux events ($|\hat{T}\hat{w}|>1$) are mostly dominated by either 3-component isotropy or 1-component anisotropy (indicated by blue and red respectively). However, for weak heat flux events ($|\hat{T}\hat{w}|<1$) the anisotropic states of the Reynolds stress tensor are dominated by 2-component anisotropy (indicated by green). This implies that the influence of the three limiting states of the Reynolds stress tensor are associated with specific heat flux events, residing within the red (1-component anisotropy), blue (3-component isotropy), and green (2-component anisotropy) regions of $\hat{T}$-$\hat{w}$ quadrant plane. We also notice from figure \ref{fig:5}a that the zones of 3-component isotropic states (blue regions) reside mainly within the warm-updraft ($C_{3 \rm c}\approx$ 0.5) and cold-downdraft ($C_{3 \rm c}\approx$ 0.4) quadrants. On the other hand, from figure \ref{fig:6}a we note that the JPDF contours between $\hat{T}$ and $\hat{w}$ depart significantly from the bivariate Gaussian distribution (see (\ref{G_f})) in highly-convective conditions. By comparing the features in figure \ref{fig:5}a with the JPDF contours in figure \ref{fig:6}a, we observe that the 1-component anisotropy zones (red regions) are associated with extremely low probability events of very high heat fluxes, located well beyond the hyperbolic hole ($|\hat{T}\hat{w}|>>1$).

However as $-\zeta$ becomes smaller, the JPDF contours become progressively close to bivariate Gaussian distribution (figures \ref{fig:6}a to \ref{fig:6}f), with the green regions (2-component anisotropy) being systematically more prominent (figures \ref{fig:5}a to \ref{fig:5}f). On the other hand, the blue regions (3-component isotropy) become systematically less visible (figures \ref{fig:5}a to \ref{fig:5}f). This is consistent with figure \ref{fig:2}a, where the anisotropic states of the Reynolds stress tensor become progressively more dominated by 2-component anisotropy as the near-neutral stability is approached. Furthermore, this is also in agreement with figure \ref{fig:3}, where highly anisotropic turbulence is associated with almost symmetrical distribution of the warm-updrafts and cold-downdrafts in near-neutral stability, due to the small values of skewness in $T^{\prime}$ (see (\ref{skew})).  

\begin{figure*}
\centering
\vspace*{0.5in}
\hspace*{-0.75in}
\includegraphics[width=1.3\textwidth]{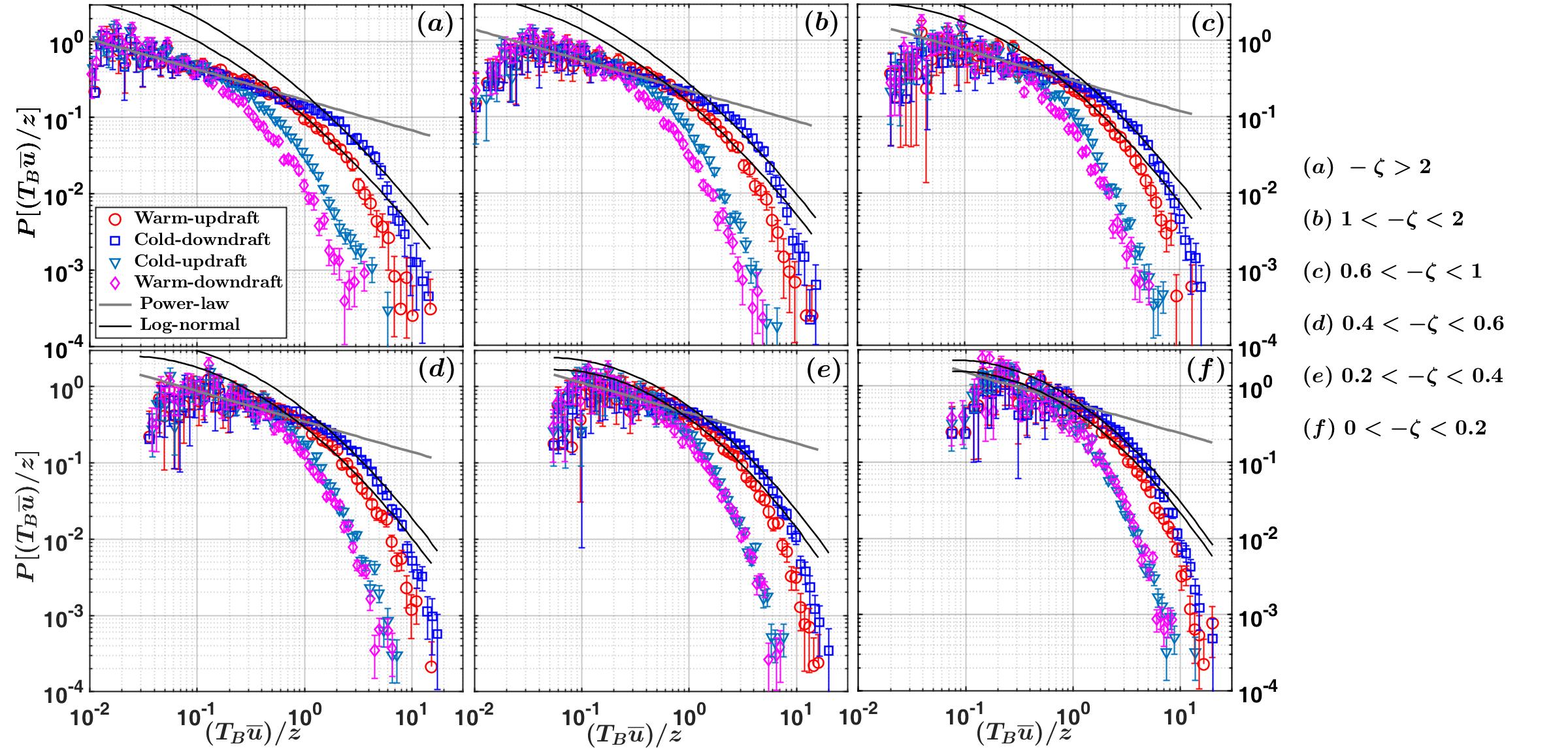}
  \caption{The persistence PDFs of the heat flux events are shown for the six different stability classes as indicated in the legend placed at the right most corner. The markers for different quadrants are explained in the legend in panel (a). A distinct power-law of exponent $-$0.4 is shown as a thick grey line in all the panels. The thick black lines correspond to the log-normal distribution (see the legend in panel (a)). The error-bars in all the panels show the existing spread between individual 30-min runs for each of the stability classes, computed as one standard deviation from the ensemble mean.}
\label{fig:7}
\end{figure*}

It is interesting to note that, the 1-component anisotropy indicated by the red regions in the $\hat{T}$-$\hat{w}$ quadrant plane does not appear to have a signature in the 30-min averaged Reynolds stress anisotropy (figure \ref{fig:2}a). This is because this anisotropic state is associated with highly-intermittent low probability events of very high heat fluxes. In addition to that, the Reynolds stress anisotropy is dominated by the 3-component isotropic state specifically for those heat flux events which reside within the blue regions of $\hat{T}$-$\hat{w}$ quadrant plane (figure \ref{fig:5}). This observation is non-trivial and this outcome would not be possible without an event based description. Since the approaches based on time-averaged statistics would predict that higher convective conditions (high heat fluxes) are associated with less anisotropic turbulence, this analysis shows that the connection between the intensity of the heat flux and turbulence anisotropy is more intricate than that. Therefore, one can ask ``\emph{whether there are any characteristic sizes of the heat flux events associated with least anisotropic turbulence}?''. However, the quadrant analysis does not give information about the time scale or size of the heat flux events. We thus focus our attention to persistence analysis to investigate the anisotropic states of the Reynolds stress tensor associated with the streamwise sizes of the heat flux events.

\subsection{Persistence analysis of Reynolds stress anisotropy}\label{results3}
We employ persistence analysis to characterize the streamwise sizes of the heat flux events from each quadrant of $T^{\prime}$-$w^{\prime}$. This is achieved by converting the persistent time $T_{B}$ to a streamwise length $T_{B}\overline{u}$ from Taylor's hypothesis. We begin with discussing the persistence PDFs to highlight the physical characteristics of these heat flux events and the aspect of non-Gaussianity. Along with that, we also investigate the anisotropic states of the Reynolds stress tensor associated with these heat flux events of different sizes. The spread in the averaged plots is shown as the error-bars, computed as one standard deviation from the ensemble mean for a particular stability class.

\subsubsection{Persistence PDFs of heat flux events}\label{results3_1}
Figures \ref{fig:7}a--f show the PDFs of the normalized streamwise sizes ($(T_{B}\overline{u})/z$) for the heat flux events occurring in each quadrant of $T^{\prime}$-$w^{\prime}$, corresponding to the six different stability classes as outlined in table \ref{tab:2}. We choose to normalize the streamwise sizes by $z$, under the assumption that these heat flux events are associated with the thermal plumes which grow linearly with height \citep{tennekes1972first}. The associated histograms of $(T_{B}\overline{u})/z$ for the heat flux events from each quadrant are also shown in figures \ref{fig:16}a--f (see appendix \ref{appA}). Typically, for the warm-updraft and cold-downdraft quadrants we encounter 100-200 number of heat flux events corresponding to the large sizes $(T_{B}\overline{u})/z>4$.

\begin{figure*}
\centering
\vspace*{0.5in}
\hspace*{-0.75in}
\includegraphics[width=1.3\textwidth]{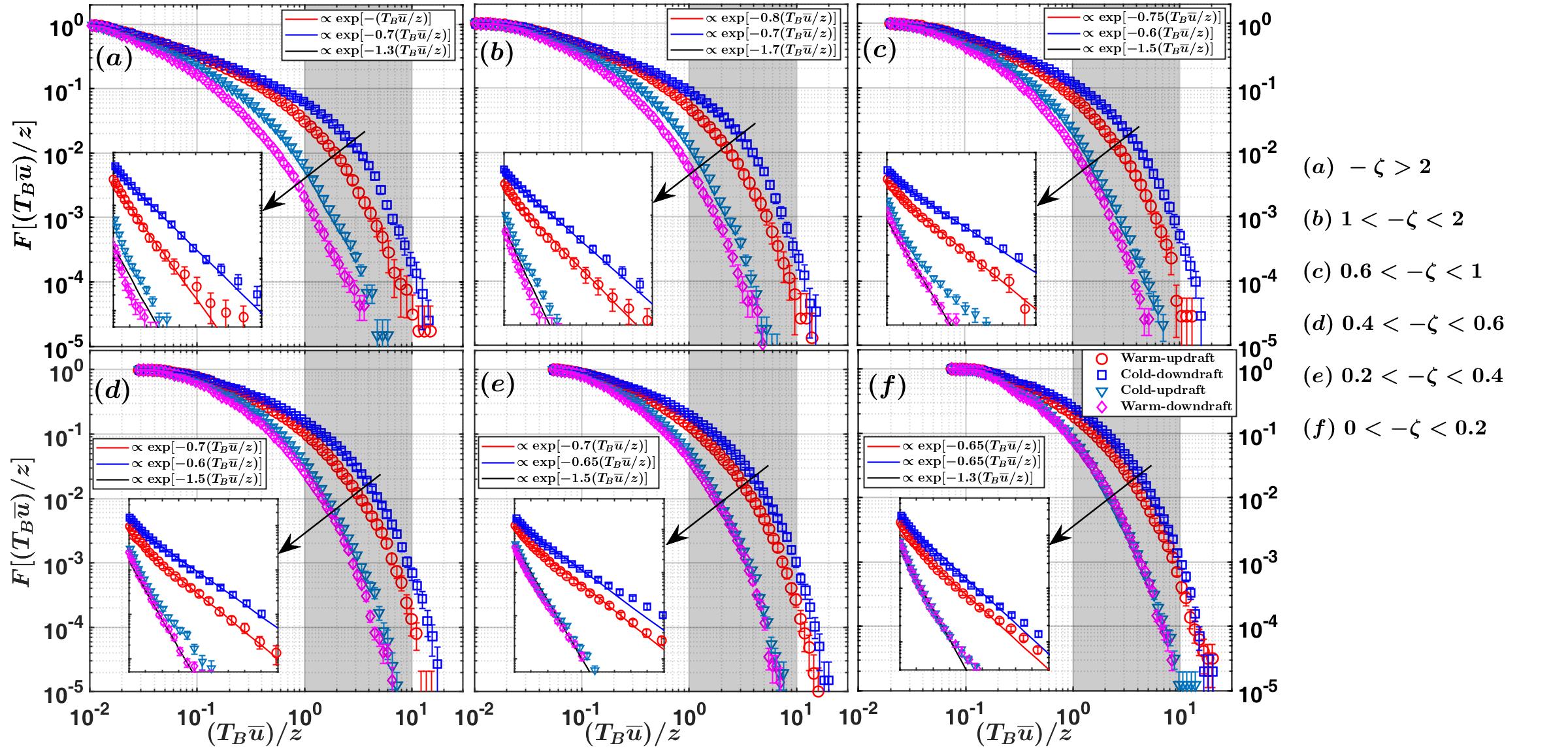}
  \caption{Same as in figure \ref{fig:7}, but the CDFs are shown instead of the PDFs. In each panel, the inset shows the enlarged area between 1 $\leq (T_{B}\overline{u})/z \leq$ 10 (grey shaded region), where the CDFs are plotted on the log-linear axes to indicate the exponential decay (Poisson process) as a straight line. The markers are explained in the legend in panel (f). The equations related to the exponential decay are shown in the legend of each panel.}
\label{fig:8}
\end{figure*}

The most distinct feature we notice from the highly-convective ($-\zeta>$ 2) stability class (figure \ref{fig:7}a) is that the persistence PDFs of the warm-updraft and cold-downdraft events collapse with a power-law of an exponent $-$0.4,
\begin{equation}
P[(T_{B}\overline{u})/z] \propto [(T_{B}\overline{u})/z]^{-0.4},
\label{pl}
\end{equation}
which approximately extends up to $(T_{B}\overline{u})/z \approx $ 1. A similar power-law was reported by \citet{chamecki2013persistence} for the persistent PDFs of $u$ and $w$ fluctuations smaller than the integral time scale in a plant canopy. Apart from that, \citet{yee1993statistical} and \citet{katul1994conditional} also documented a power-law behaviour in the PDFs of the small sizes of the heat flux bursts from an unstable ASL, although the exponent they found was closer to $-1.4$. Additionally, we also note that the PDFs of $(T_{B}\overline{u})/z$ for the counter-gradient events are in close agreement with the PDFs of the down-gradient events for small values of $(T_{B}\overline{u})/z<$ 0.2. Beyond those sizes, the PDFs of the counter-gradient events drop faster than the down-gradient events. This implies that these counter-gradient events have a statistical tendency to occur in smaller sizes and do not persist for a long time. This is in agreement with the simulations of \citet{dong2017coherent}, where they found that the PDFs of counter-gradient and down-gradient momentum events of small sizes agreed with each other and diverged for larger sizes (see their figure 5b). 

However, this power-law segment systematically disappears as we approach the near-neutral stability (0 $<-\zeta<$ 0.2) and gets replaced by a log-normal distribution (figure \ref{fig:7}f). According to \citet{cava2009effects}, the log-normal distribution can be expressed as,
\begin{equation}
P[(T_{B}\overline{u})/z] \propto \exp(a{\alpha}^2+b\alpha+c),
\label{ln}
\end{equation}
where $\alpha=\log[(T_{B}\overline{u})/z]$, and $a$, $b$, $c$ are related to the variance ($\sigma^2$) and mean ($\mu$) of $\log[(T_{B}\overline{u})/z]$ as,
\begin{equation}
\begin{split}
a=-1/(2\sigma^2) \\
b=-1+2 \mu a \\
c=-\log(\sqrt{2\pi}\sigma)-\mu^2 a
\end{split}.
\end{equation}
This is broadly in agreement with \citet{sreenivasan2006clustering}, where they commented that for an active scalar such as temperature in highly-convective turbulence, the PDFs of the inter-pulse periods followed a power-law. Conversely in a shear-driven turbulence when the temperature behaved more like a passive scalar, the PDFs followed a log-normal distribution.

For $(T_{B}\overline{u})/z>$ 1, the PDFs of the warm-updraft and cold-downdraft events significantly differ from each other in the highly-convective case ($-\zeta>$ 2, figure \ref{fig:7}a). However, they systematically agree with each other as the near-neutral stability (0 $<-\zeta<$ 0.2) is approached (figures \ref{fig:7}a to \ref{fig:7}f). As we will show later, this is related to the asymmetry in the distributions of the warm-updraft and cold-downdraft events due to strong non-Gaussianity in temperature fluctuations in a highly-convective surface layer. As discussed by \citet{chamecki2013persistence}, these large values of $(T_{B}\overline{u})/z$ are exponentially distributed according to a Poisson type process which could be studied by considering the cumulative distribution functions (CDFs) of $(T_{B}\overline{u})/z$. In general, the CDFs are comparatively smoother than the PDFs, thus yielding a more robust fit for the exponential distribution. The CDF ($F[(T_{B}\overline{u})/z]$) is defined as,
\begin{equation}
F\Big[(T_{B}\overline{u})/z\Big]=\int_{{[(T_{B}\overline{u})/z]}_{\rm max}}^{[(T_{B}\overline{u})/z]} P\Big[(T_{B}\overline{u})/z\Big] \, d \log{\Big[(T_{B}\overline{u})/z\Big]}.
\label{P_f_3}
\end{equation}

\begin{figure*}
\centering
\vspace*{0.5in}
\hspace*{-2.3in}
\includegraphics[width=1.8\textwidth]{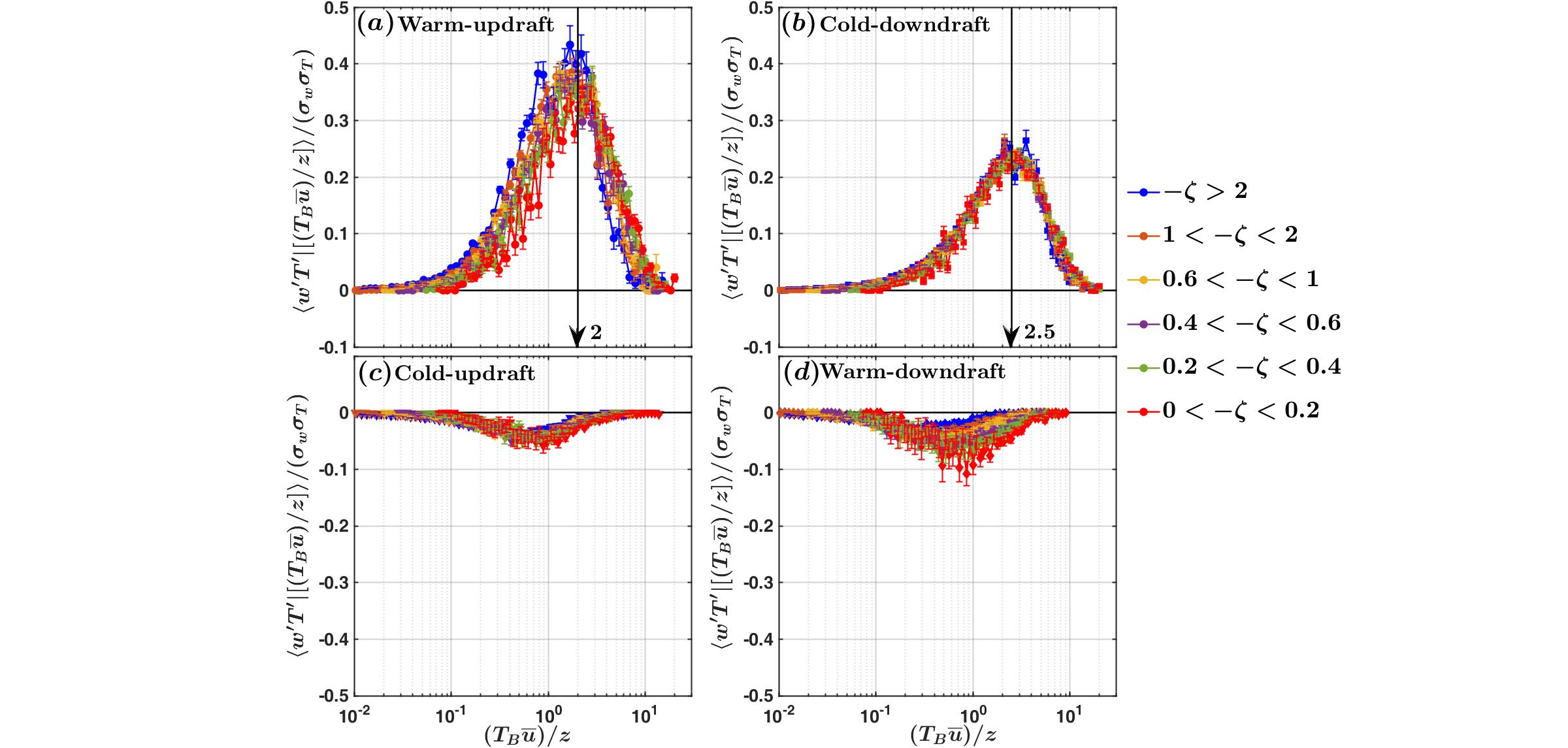}
  \caption{The heat flux distributions (see (\ref{F_1})) are plotted against the normalized sizes $(T_{B}\overline{u})/z$ of the heat flux events corresponding to (a) warm-updraft, (b) cold-downdraft, (c) cold-updraft, and (d) warm-downdraft quadrants. In the top panel, the thick black arrows indicate the collapsed position of the peaks of the heat flux distribution associated with the warm-updrafts and cold-downdrafts. The different colours represent the six different stability classes as indicated in the legend placed at the right most corner.}
\label{fig:9}
\end{figure*}

Figure \ref{fig:8} shows the CDFs of the heat flux events from the four quadrants of $T^{\prime}$-$w^{\prime}$. We note that the power-law region is not seen clearly in the CDFs as they converge to 1 for the small sizes $(T_{B}\overline{u})/z<$ 1. For the large sizes (1 $\leq (T_{B}\overline{u})/z \leq$ 10), we plot the CDFs in a log-linear coordinate system (see the insets in figure \ref{fig:8}) such that the exponential decay,
\begin{equation}
F\Big[\frac{(T_{B}\overline{u})}{z}\Big] \propto \exp \Big[-k\frac{(T_{B}\overline{u})}{z}\Big],
\label{P_f_4}
\end{equation}
in such plots would appear as a straight line with the slope of $-k$. From the insets in figure \ref{fig:8}, we notice that for larger values of $(T_{B}\overline{u})/z$, $F[(T_{B}\overline{u})/z]$ indeed decays exponentially according to (\ref{P_f_4}). We also find that the slopes corresponding to the warm-updraft and cold-downdraft events are significantly different from each other ($k=1.3$ and $k=0.7$ respectively) for highly-convective stability (figure \ref{fig:8}a). However, these two slopes become systematically close to each other as the near-neutral stability is approached ($k=0.65$, figure \ref{fig:8}f). On the other hand, with stability no appreciable change in the slope is observed for the counter-gradient events. This difference in the slopes for warm-updraft and cold-downdraft events is linked to the non-Gaussianity in temperature fluctuations in a highly convective surface layer. \citet{sreenivasan1983zero} mentioned that the long intervals (large $(T_{B}\overline{u})/z$) are a consequence of large-scale structures passing the sensor and the short intervals (small $(T_{B}\overline{u})/z$) are a consequence of the nibbling small-scale motions superposed on the large-scale structures. From that perspective, we expect that the non-Gaussian characteristics of the warm-updraft and cold-downdraft events might be related to the large-scale structures. 

To summarize, from figures \ref{fig:7}--\ref{fig:8} we have observed that the warm-updraft and cold-downdraft events having sizes $(T_{B}\overline{u})/z<$ 1 are scale-invariant owing to a power-law dependency in the highly-convective stability. This scale-invariant property disappears systematically as the near-neutral stability is approached. Apart from that, the effect of non-Gaussianity (Gaussianity) appears mostly at the sizes $(T_{B}\overline{u})/z>$ 1 in a highly (weakly) convective surface layer, possibly associated with the large-scale structures \citep{sreenivasan1983zero}. We will revisit this while investigating the linkage between the persistence PDFs and the degree of isotropy of the Reynolds stress tensor in \S\ref{results3_4}. Next we discuss the anisotropy characteristics of the Reynolds stress tensor associated with these heat flux events of different sizes.

\subsubsection{The degree of isotropy, heat, and momentum flux distributions}\label{results3_2}
We begin with discussing the amount of heat flux associated with the normalized streamwise sizes $(T_{B}\overline{u})/z$ (see (\ref{F_1})), corresponding to the six different stability classes. From figures \ref{fig:9}a--b, we note that the $z$-scaling of the streamwise sizes of the heat flux events collapses the scaled heat flux peak positions at $(T_{B}\overline{u})/z \approx 2$ and $(T_{B}\overline{u})/z \approx 2.5$ for the warm-updraft and cold-downdraft quadrants respectively. We also observe that the heat flux events from the counter-gradient quadrants contribute insignificantly to the total heat flux (figures \ref{fig:9}c--d). This result is in agreement with the heat flux fractions shown in figure S3 (supplementary material). To infer whether the least anisotropic turbulence is associated with the peak positions of the heat flux, we investigate the distributions of the degree of isotropy ($C_{3 \rm c}$, see (\ref{bm4_1})) associated with these events.

\begin{figure*}
\centering
\vspace*{0.5in}
\hspace*{-2.7in}
\includegraphics[width=1.7\textwidth]{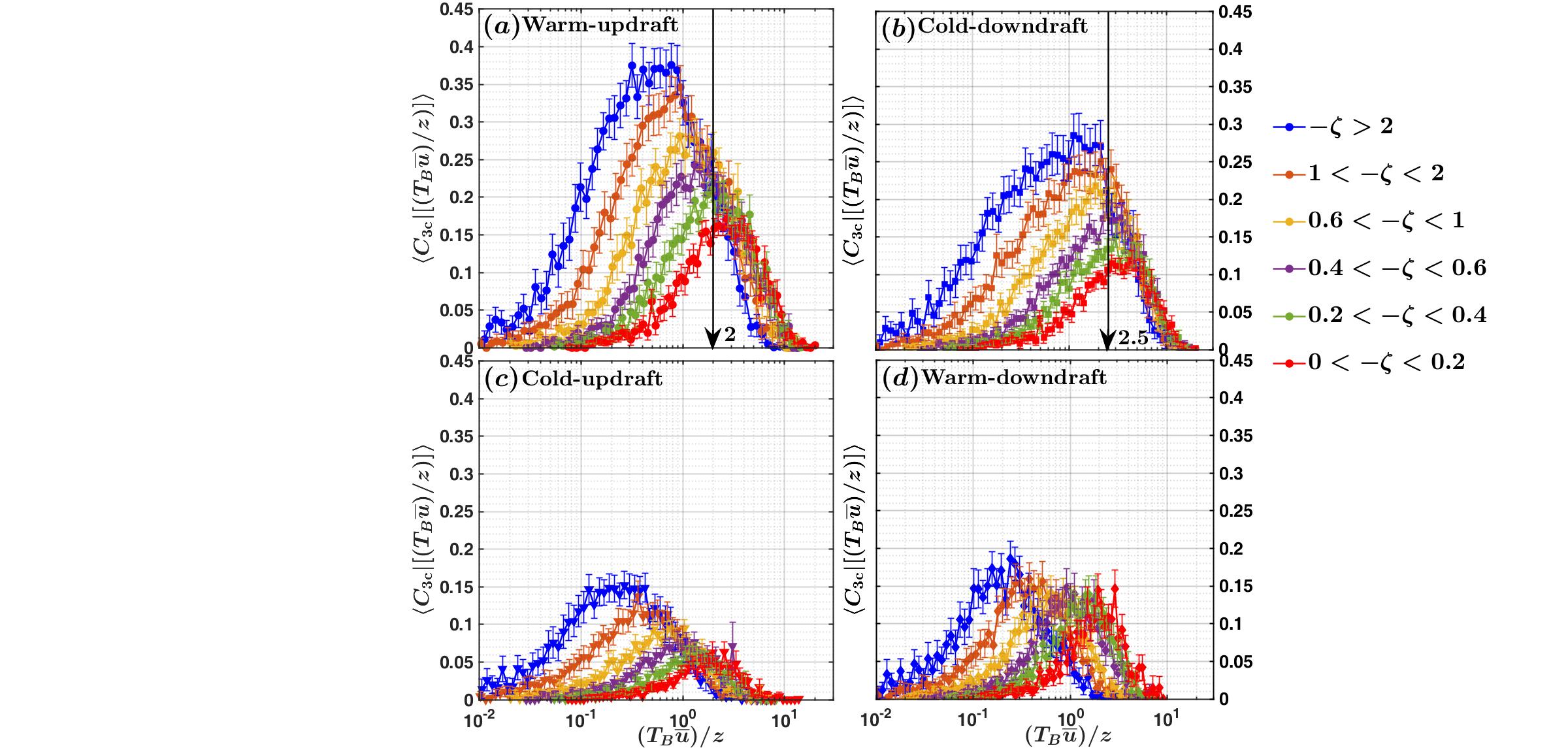}
  \caption{The distributions of the degree of isotropy ($C_{3 \rm c}$) are plotted against the normalized sizes $(T_{B}\overline{u})/z$ of the heat flux events corresponding to (a) warm-updraft, (b) cold-downdraft, (c) cold-updraft, and (d) warm-downdraft quadrants. In the top panel, the thick black arrows indicate the collapsed position of the peaks of the heat flux distribution associated with the warm-updrafts and cold-downdrafts (figure \ref{fig:9}).}
\label{fig:10}
\end{figure*}

From figure \ref{fig:10}, we note that the down-gradient heat flux events corresponding to warm-updraft and cold-downdraft quadrants are associated with larger values of $C_{3 \rm c}$, compared to the counter-gradient heat flux events from cold-updraft and warm-downdraft quadrants. Therefore, we may infer that the counter-gradient events which carry significantly less heat flux are associated with more anisotropic turbulence than the warm-updraft and cold-downdraft events. Apart from that, we observe that there is a critical size of warm-updraft and cold-downdraft events associated with the maximum value of $C_{3 \rm c}$ and this critical size is larger for the cold-downdrafts compared to the warm-updrafts. Also, the maximum value of $C_{3 \rm c}$ decreases systematically as the near-neutral stability is approached. This is in agreement with our previous observations for the averaged flow, where the degree of isotropy systematically decreased from highly-convective to near-neutral stability (figures \ref{fig:2} and \ref{fig:3}). Moreover, the heat flux events corresponding to warm-updrafts are associated with relatively less anisotropic turbulence than the cold-downdrafts as the values of $C_{3 \rm c}$ are larger in general. However, the peak positions of $C_{3 \rm c}$ associated with warm-updraft and cold-downdraft events do not match with the peak positions of the heat flux distributions (figures \ref{fig:10}a--b). This mismatch is more apparent for the warm-updraft events than the cold-downdrafts. The deviation in the peak positions of $C_{3 \rm c}$ and heat flux distributions complements the results from the quadrant analysis (figures \ref{fig:5} and \ref{fig:6}), where we found that the large heat flux events do not necessarily relate to the least anisotropic turbulence.

\begin{figure*}
\centering
\vspace*{0.5in}
\hspace*{-2in}
\includegraphics[width=1.5\textwidth]{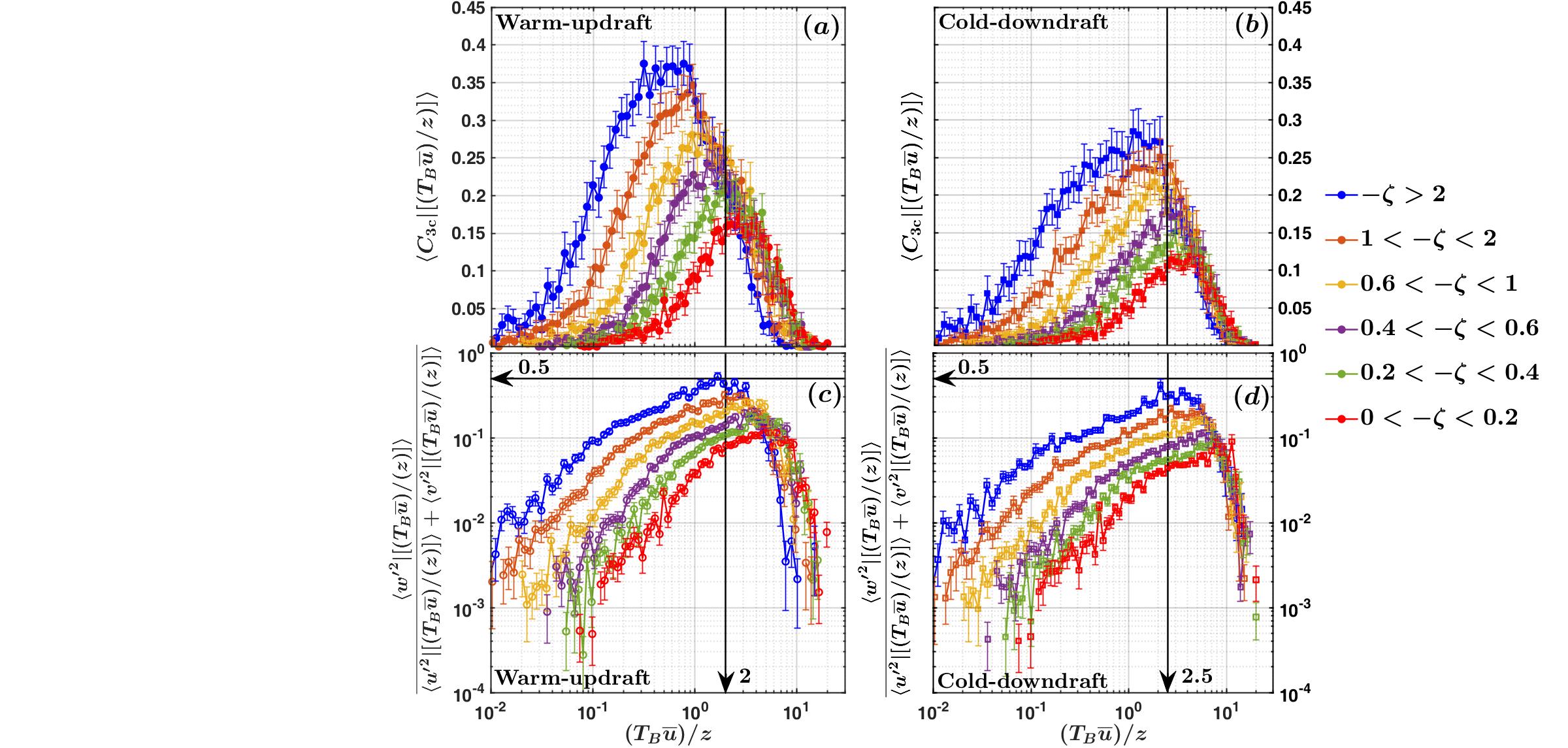}
  \caption{The distributions of the degree of isotropy are plotted against the normalized sizes $(T_{B}\overline{u})/z$ of the heat flux events from (a) warm-updraft and (b) cold-downdraft quadrants, as shown in the top panel. In the bottom panel, the ratio between the vertical and horizontal velocity variances are plotted against the normalized sizes $(T_{B}\overline{u})/z$ of the heat flux events from (c) warm-updraft and (d) cold-downdraft quadrants. The horizontal black arrows in panels (c) and (d) indicate the ratio 0.5.}
\label{fig:11}
\end{figure*}

From the definition of isotropy \citep{konozsy2019new}, there are two possible reasons contributing to the Reynolds stress anisotropy associated with the sizes of the warm-updraft and cold-downdraft events, such as:
\begin{enumerate}
\item The amplitudes of the horizontal velocity fluctuations exceed the vertical velocity fluctuations. \item The vertical velocity fluctuations contribute substantially to the upward or downward transport of streamwise momentum. 
\end{enumerate}
To investigate the first of the two aforementioned reasons, in figure \ref{fig:11} we show the ratios of the vertical and horizontal velocity variances associated with the normalized sizes $(T_{B}\overline{u})/z$ of the warm-updraft and cold-downdraft events. The velocity variances associated with $(T_{B}\overline{u})/z$ from each quadrant of $T^{\prime}$-$w^{\prime}$ are defined similarly as in \S\ref{Prob_flux_persistence}, such that:
\begin{align}
\begin{split}
\Big \langle {x^{\prime}}^2 \vert\Big[(T_{B}\overline{u}/z)_{\rm bin}\{m\}< (T_{B}\overline{u}/z)< (T_{B}\overline{u}/z)_{\rm bin}\{m\}+d \log{(T_{B}\overline{u}/z)}\Big] \Big \rangle &=
\\
\frac{\sum{{x^{\prime}}^2}}{N \times d \log{(T_{B}\overline{u}/z)}}, \  (x=u,v,w)
\end{split}
\label{Vf_1}
\end{align}
where the symbols carry their same meaning. Since in isotropic turbulence the three velocity variances in $x$, $y$, and $z$ direction are equal to each other, it follows that, 
\begin{equation}
\frac{\langle {w^{\prime}}^2 \vert [(T_{B}\overline{u})/z] \rangle}{\langle {u^{\prime}}^2 \vert [(T_{B}\overline{u})/z] \rangle + \langle {v^{\prime}}^2 \vert [(T_{B}\overline{u})/z] \rangle}=0.5.
\label{var_ratio}
\end{equation}

From figures \ref{fig:11}c--d, we note that for the sizes of the warm-updraft and cold-downdraft events $(T_{B}\overline{u})/z < 2$ and $(T_{B}\overline{u})/z < 2.5$ respectively, the horizontal velocity variances dominate, since the variance ratio is smaller than 0.5. At the peak position of the heat flux, the variance ratio indeed reach closer to 0.5 for the highly-convective stability and then systematically decrease as the near-neutral stability is approached (figures \ref{fig:11}c--d). However, for the highly-convective stability the maximum in the degree of isotropy associated with these events occurs at relatively smaller sizes than the peak positions of the heat flux distribution (figures \ref{fig:11}a--b). Therefore, the disagreement in the peak positions of heat flux and degree of isotropy might be related to the second reason associated with momentum transport.

\begin{figure*}
\centering
\vspace*{0.5in}
\hspace*{-2in}
\includegraphics[width=1.5\textwidth]{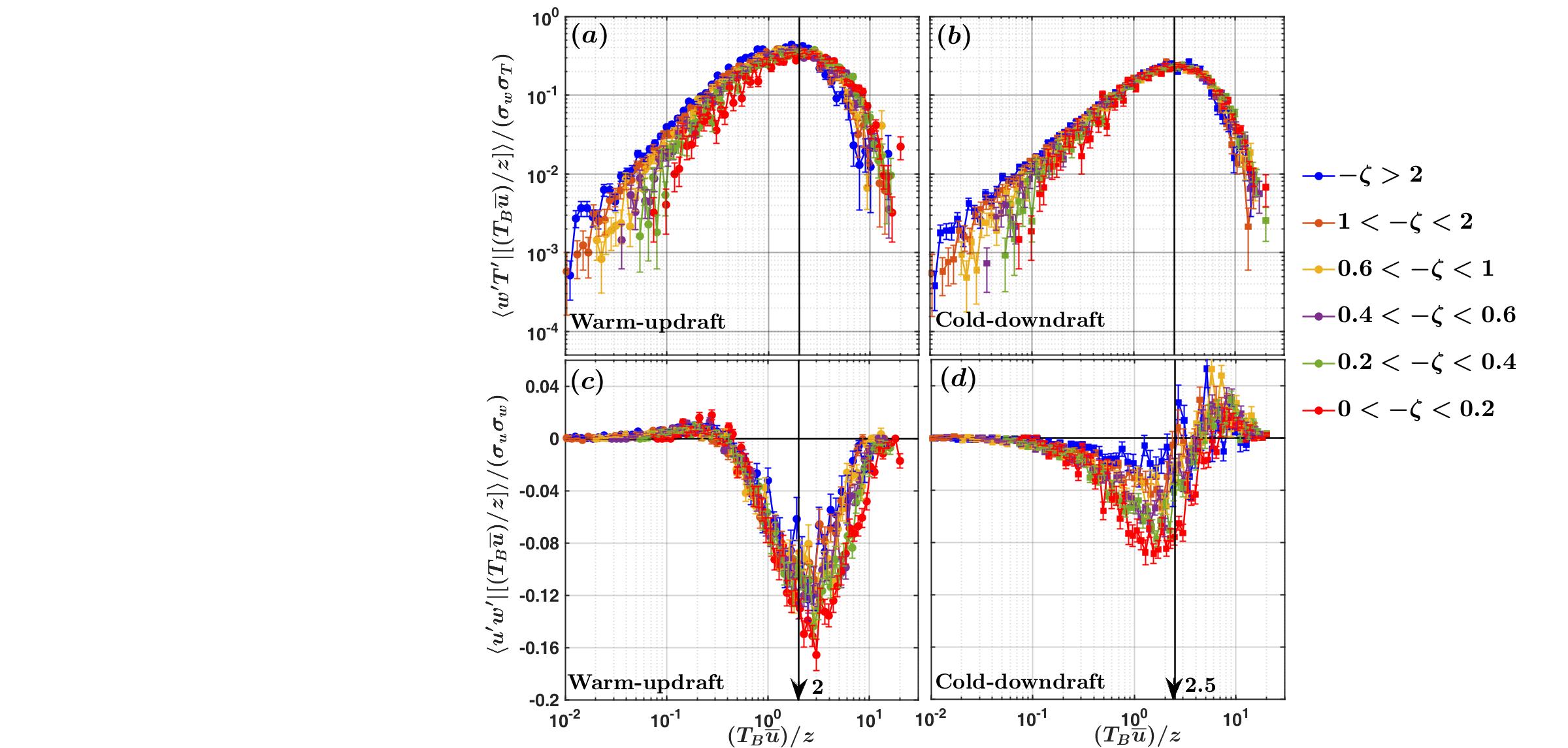}
  \caption{The scaled heat and momentum flux distribution plotted against the normalized streamwise sizes $(T_{B}\overline{u})/z$ of the heat flux events corresponding to the warm-updraft (a, c) and cold-downdraft (b, d) quadrants.}
\label{fig:12}
\end{figure*}

Figure \ref{fig:12} shows the distributions of the heat and momentum fluxes associated with the warm-updraft and cold-downdraft events of different sizes, $(T_{B}\overline{u})/z$ (see (\ref{F_1})). It is clear that the heat flux peak position associated with warm-updraft events ($(T_{B}\overline{u})/z \approx 2$), corresponds to significant amount of down-gradient momentum in a highly-convective surface layer (figures \ref{fig:12}a and c). However, for the peak position $(T_{B}\overline{u})/z \approx 2.5$ associated with the cold-downdraft events, the momentum transport is rather erratic in nature (figures \ref{fig:12}b and d). The association of highly erratic momentum transport with the cold-downdrafts has been observed in the numerical simulations of \citet{li2018implications} and in the observations of \citet{chowdhuri2019evaluation}. \citet{salesky2018buoyancy} interpreted this as, under highly-convective conditions, the small-scale turbulence is excited in the updraft regions and suppressed in downdraft regions, leading to intermittent periods of small-scale excitation in the momentum fluxes.

Summarizing these observations, we note that there is a characteristic size of the warm-updraft and cold-downdraft events associated with least anisotropic turbulence, which does not scale with $z$. On the other hand, the sizes of the warm-updraft and cold-downdraft events which carry the maximum heat are found to scale with $z$. The mismatch in the peak positions of heat flux and degree of isotropy is related to the fact that the warm-updraft events which carry the maximum amount of heat are also associated with significant down-gradient momentum transport (figures \ref{fig:12}a and c). However, for the cold-downdraft events these two peak positions almost coincide (figures \ref{fig:10}b and \ref{fig:11}b). This might be related to inefficient momentum transport associated with the cold-downdraft events, unlike the warm-updrafts (figures \ref{fig:12}c and d).

So far, we have focused on the degree of isotropy ($C_{3 \rm c}$) of the Reynolds stress tensor associated with the heat flux events of different sizes, to quantify how closer the turbulence is to 3-component isotropy. However, apart from 3-component isotropy there are 1-component and 2-component anisotropic states whose dominance is described by other two coefficients $C_{1 \rm c}$ and $C_{2 \rm c}$, respectively. Therefore, we can ascertain the entire anisotropic states of the heat flux events of different sizes, by investigating the distributions of all the three coefficients such as: $C_{1 \rm c}$, $C_{2 \rm c}$, and $C_{3 \rm c}$. We focus our attention on the down-gradient events, since for these events there is an intricate relation between the heat flux intensity and the degree of isotropy.

\subsubsection{The anisotropic states of the Reynolds stress tensor}\label{results3_3}
\begin{figure*}
\centering
\vspace*{0.5in}
\hspace*{-0.9in}
\includegraphics[width=1.3\textwidth]{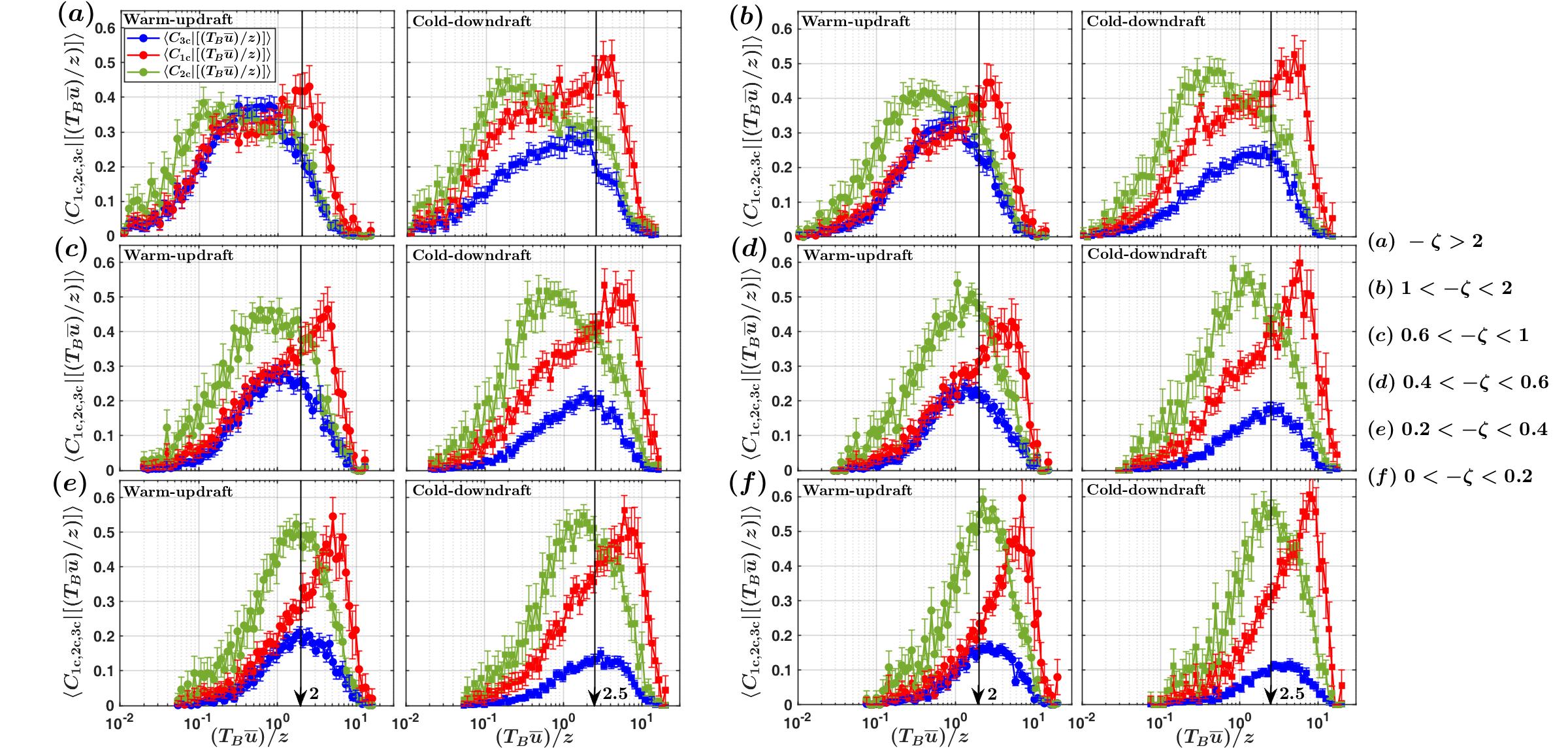}
  \caption{The three coefficients ($C_{1 \rm c}$, $C_{2 \rm c}$, and $C_{3 \rm c}$) associated with the three limiting states of the anisotropy Reynolds stress tensor are plotted against the normalized streamwise sizes $(T_{B}\overline{u})/z$ of the heat flux events corresponding to warm-updraft and cold-downdraft quadrants. The markers associated with these three coefficients are explained in the legend of panel (a). The thick black lines in all the panels indicate the collapsed position of the peaks of the heat flux distribution associated with the warm-updrafts and cold-downdrafts (figure \ref{fig:9}).}
\label{fig:13}
\end{figure*}

Figure \ref{fig:13} shows the three coefficients associated with the three limiting states of the anisotropy Reynolds stress tensor to describe the anisotropic states of the warm-updraft and cold-downdraft events (see (\ref{bm4_1})). These three coefficients describe the corresponding weights associated with each of the three limiting states of the anisotropy Reynolds stress tensor, such as: $C_{1 \rm c}$ is related to 1-component anisotropy (red lines in figure \ref{fig:13}), $C_{2 \rm c}$ is related to 2-component anisotropy (green lines in figure \ref{fig:13}), and $C_{3 \rm c}$ is related to 3-component isotropy (blue lines in figure \ref{fig:13}). Note that, from (\ref{bm4_1}) the sum of the three anisotropy coefficients should be 1 for each $(T_{B}\overline{u})/z$ values. Nevertheless, the graphs shown in figure \ref{fig:13} are ensemble averaged over all the runs from a particular stability class. Due to such ensemble averaging, the sums of the three coefficients may differ from 1 for some $(T_{B}\overline{u})/z$ values. This occurs because, some of the bins of $(T_{B}\overline{u})/z$ values may remain empty for some specific runs which construct the ensemble.   

For the highly-convective stability (figure \ref{fig:13}a), we note that for the warm-updraft events the maximum in $C_{3 \rm c}$ is located at $(T_{B}\overline{u})/z \approx$ 0.5. Moreover, we also observe that for the warm-updraft events smaller than this size ($(T_{B}\overline{u})/z <$ 0.5), the values of the coefficient $C_{2 \rm c}$ exceed the other two coefficients. On the other hand, for the sizes of warm-updraft events larger than $(T_{B}\overline{u})/z >$ 0.5, the coefficient $C_{1 \rm c}$ is the largest amongst the three and its peak position coincides with the heat flux peak position. This implies that the anisotropic states of the Reynolds stress tensor associated with the warm-updraft events smaller (larger) than the critical size $(T_{B}\overline{u})/z \approx$ 0.5 are dominated by 2-component (1-component) anisotropy.

However, for the cold-downdraft events from highly-convective stability (figure \ref{fig:13}a), the coefficients $C_{1 \rm c}$ and $C_{2 \rm c}$ dominate over $C_{3 \rm c}$ for all the sizes, albeit $C_{2 \rm c}$ being the largest for sizes $(T_{B}\overline{u})/z <$ 1. For sizes $(T_{B}\overline{u})/z >$ 1, the coefficient $C_{1 \rm c}$ is the largest and its peak position almost coincides with the maximum heat flux associated with cold-downdraft events. Interestingly enough, we also find that as the near-neutral stability is approached (figures \ref{fig:13}a to \ref{fig:13}f), the coefficient $C_{2 \rm c}$ systematically becomes the largest amongst the three for most of the sizes of warm-updraft and cold-downdraft events. Nonetheless, at larger sizes (in the order of the heat flux peak positions) there remains a tendency for the coefficient $C_{1 \rm c}$ to dominate the anisotropic state of the Reynolds stress tensor.

The results from figure \ref{fig:13} are in accordance with the results from quadrant analysis in figures \ref{fig:5} and \ref{fig:6}. For the highly-convective stability, we note that the blue regions (dominated by 3-component isotropy) in the anisotropy contour maps (figure \ref{fig:5}a) broadly correspond to the critical sizes of warm-updrafts and cold-downdrafts ($(T_{B}\overline{u})/z \approx$ 0.5 and $(T_{B}\overline{u})/z \approx$ 1) where $C_{3 \rm c}$ values are maximum. For the sizes smaller (larger) than this, the anisotropic states of the Reynolds stress tensor are dominated by 2-component (1-component) anisotropy, shown as green (red) regions in figure \ref{fig:5}a. However, for the near-neutral stability (figures \ref{fig:5}f and \ref{fig:13}f) the dominance of 2-component anisotropy is associated with almost all the sizes of the warm-updraft and cold-downdraft events, except at the larger sizes where there is a signature of 1-component anisotropy.

Hitherto, from analyses presented in figures \ref{fig:10}--\ref{fig:13} we have found that the least anisotropic turbulence is associated with particular sizes of warm-updraft and cold-downdraft events. These sizes do not exactly correspond to the peak positions of the heat flux distribution and also do not scale with $z$. With stability (highly-convective to near-neutral) this particular size changes from $(T_{B}\overline{u})/z \approx$ 0.5 to $(T_{B}\overline{u})/z \approx$ 2 for the warm-updraft events and from $(T_{B}\overline{u})/z \approx$ 1 to $(T_{B}\overline{u})/z \approx$ 3 for the cold-downdraft events. From the persistence PDFs and CDFs of these events presented in figures \ref{fig:7} and \ref{fig:8}, we have noted that there is a power-law behaviour associated with sizes $(T_{B}\overline{u})/z <$ 1, followed by an exponential decay (Poisson type process) for sizes $(T_{B}\overline{u})/z >$ 1. In the following section, we present results to investigate whether there is any correspondence between these PDFs and the critical sizes of warm-updraft and cold-downdraft events associated with least anisotropic turbulence.

\begin{figure*}
\centering
\vspace*{0.5in}
\hspace*{-0.85in}
\includegraphics[width=1.3\textwidth]{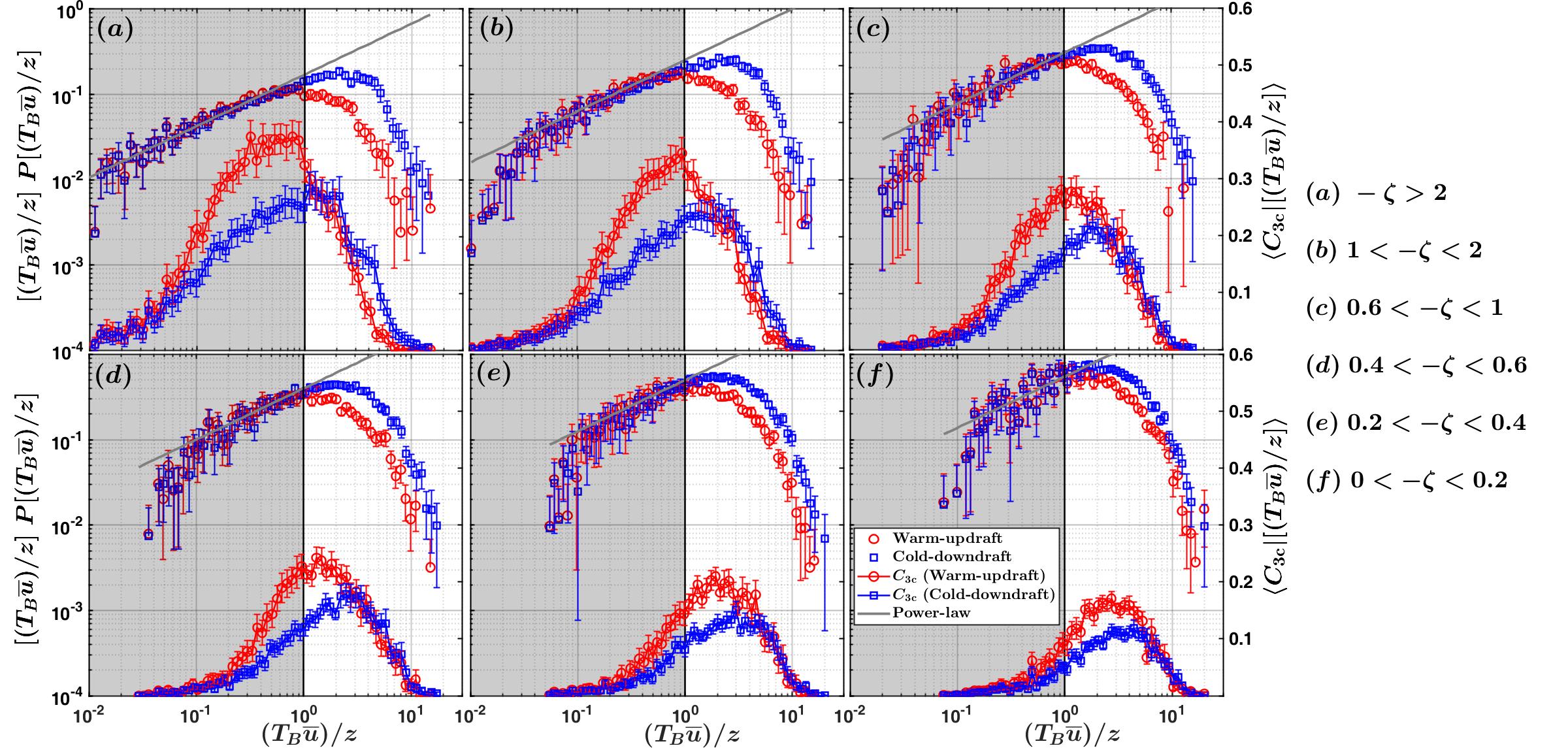}
  \caption{The log-log plots of the premultiplied PDFs of $(T_{B}\overline{u})/z$ (see (\ref{tf_1})) corresponding to the heat flux events from the warm-updraft and cold-downdraft quadrants are shown for the six different stability classes, as indicated in the legend placed at the right-most corner. In all the panels, the right hand side of the $y$ axis is linear and used to represent the distribution of the degree of isotropy ($\langle C_{3 \rm c}\vert [(T_{B}\overline{u})/z] \rangle$) associated with the warm-updraft and cold-downdraft events. The thick grey line shows the same power-law as in figure \ref{fig:7}, but owing to premultiplication the exponent changed to $+$0.6. The grey shaded region represents $(T_{B}\overline{u})/z<$ 1, and the thick black line denotes the value of 1. The markers are explained in the legend in panel (f).}
\label{fig:14}
\end{figure*}

\begin{figure*}
\centering
\vspace*{0.5in}
\hspace*{-1.65in}
\includegraphics[width=1.5\textwidth]{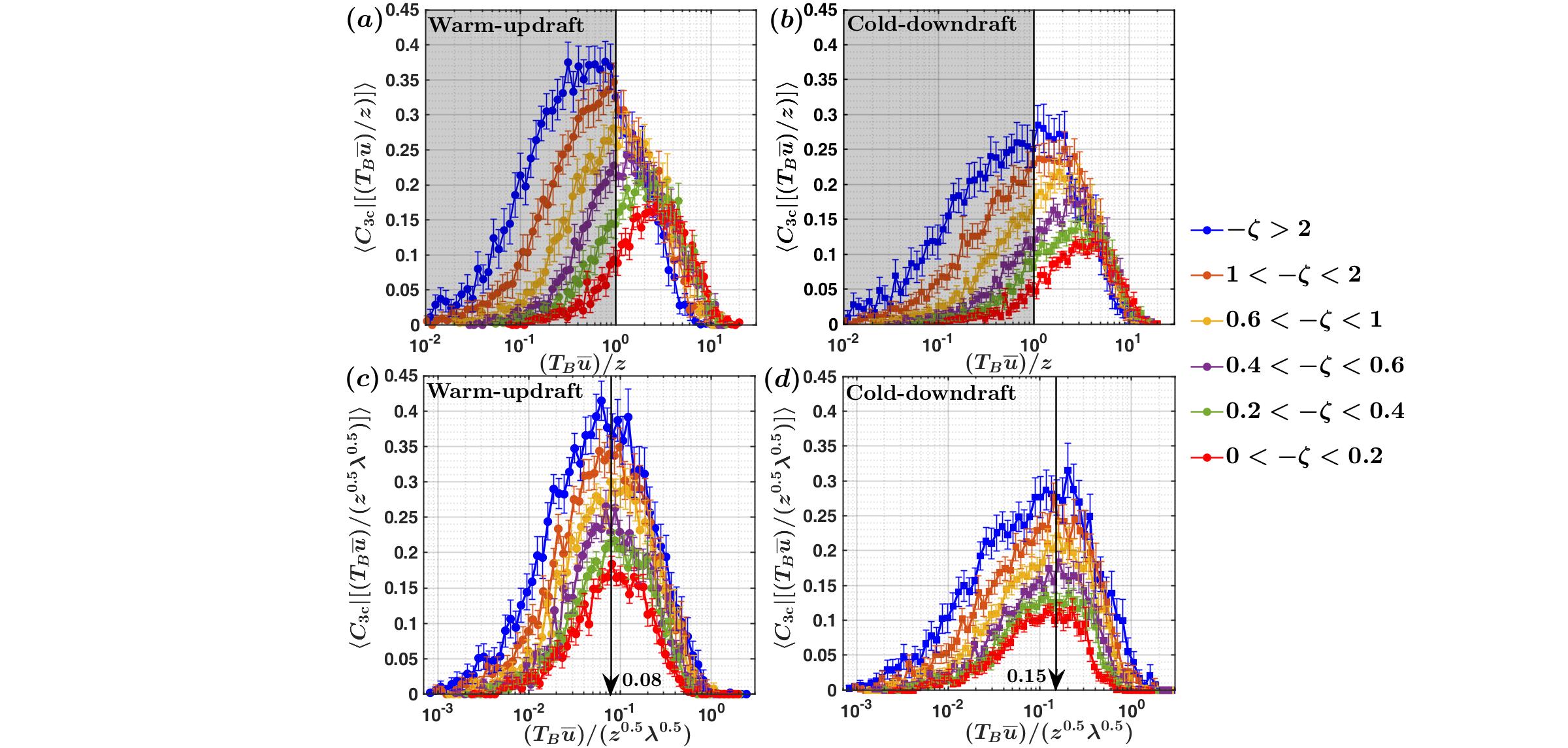}
  \caption{The distributions of the degree of isotropy ($C_{3 \rm c}$) for the $z$- and mixed-length scaled sizes of the heat flux events from warm-updraft and cold-downdraft quadrants ($(T_{B}\overline{u})/z$ and $(T_{B}\overline{u})/(z^{0.5}\lambda^{0.5})$) are shown in the top and bottom panels, respectively. The scale $\lambda$ is the large-eddy length scale obtained from (\ref{model}). In the top panel, the grey shaded region represents $(T_{B}\overline{u})/z<$ 1 and the thick black line denotes the value of 1. In the bottom panel, the black arrows indicate the peak positions of $C_{3 \rm c}$ corresponding to the mixed-length scaled sizes of the heat flux events from warm-updraft and cold-downdraft quadrants.}
\label{fig:15}
\end{figure*}

\subsubsection{The linkage between degree of isotropy and persistence PDFs}\label{results3_4}
Before discussing anisotropy, to highlight non-Gaussianity we convert the PDFs in figure \ref{fig:7} to a distribution about the time fractions ($T_{f}$) spent in each quadrant of $T^{\prime}$-$w^{\prime}$, by presenting the same in a premultiplied form. If from a particular quadrant of $T^{\prime}$-$w^{\prime}$, $N_{\rm tot}$ number of blocks are being detected, with each $N_{i}$-th block containing $n_{i}$ number of points, then we can write, 
\begin{equation}
\sum\limits_{i=1}^{N_{\rm tot}}{N_{i}n_{i}} \propto T_{f},
\label{tf_1_1}
\end{equation}
where $T_{f}$ is the time fraction spent in that particular quadrant. Since, the probability of finding a block containing $n_{i}$ number of points is $N_{i}/N_{\rm tot}$, from (\ref{P_f_1}) we can write,
\begin{equation*}
N_{i}\propto \Big(P[(T_{B}\overline{u})/z] \hspace{0.15em} d \log{[(T_{B}\overline{u})/z]}\Big) \ \textrm{and} \ n_{i}\propto (T_{B}\overline{u})/z.
\end{equation*}
Therefore (\ref{tf_1_1}) can be expressed as,
\begin{equation}
\int_{(\frac{T_{B}\overline{u}}{z})_{\rm min}}^{(\frac{T_{B}\overline{u}}{z})_{\rm max}} (\frac{T_{B}\overline{u}}{z}) \ P[(\frac{T_{B}\overline{u}}{z})] \ d \log{(\frac{T_{B}\overline{u}}{z})} \propto T_{f}.
\label{tf_1}
\end{equation}
From (\ref{tf_1}) we can also write,
\begin{equation}
\int_{(\frac{T_{B}\overline{u}}{z})_{\rm min}}^{(\frac{T_{B}\overline{u}}{z})_{\rm max}} \Big( \Big [\frac{T_{B}\overline{u}}{z} P(\frac{T_{B}\overline{u}}{z}) \Big]_{\rm III}-\Big [\frac{T_{B}\overline{u}}{z} P(\frac{T_{B}\overline{u}}{z}) \Big]_{\rm I}\Big) \ d \log{(\frac{T_{B}\overline{u}}{z})} \propto \Delta T_{f},
\label{tf_2}
\end{equation}
where the subscripts I and III refer to the warm-updraft and cold-downdraft quadrants, and $\Delta T_{f}$ is the difference in the time fractions spent in those quadrants. From (\ref{skew}) we know that $\Delta T_{f} \approx \overline{{T^{\prime}}^3}/\sigma_{T}^3$, given the assumption that the time-fractions spent in the counter-gradient quadrants could be neglected. Since from figure \ref{fig:7} we have noticed that the persistence PDFs of the counter-gradient events decrease faster than the down-gradient events for the large sizes, we may rewrite (\ref{tf_2}) as,

\begin{equation}
\int_{(\frac{T_{B}\overline{u}}{z})_{\rm min}}^{(\frac{T_{B}\overline{u}}{z})_{\rm max}} \Big( \Big [\frac{T_{B}\overline{u}}{z} P(\frac{T_{B}\overline{u}}{z}) \Big]_{\rm III}-\Big [\frac{T_{B}\overline{u}}{z} P(\frac{T_{B}\overline{u}}{z}) \Big]_{\rm I} \Big) \ d \log{(\frac{T_{B}\overline{u}}{z})} \propto \frac{\overline{{T^{\prime}}^3}}{\sigma_{T}^3}.
\label{tf_3}
\end{equation}
 
Figures \ref{fig:14}a--f show the premultiplied PDFs of $(T_{B}\overline{u})/z$ corresponding to the warm-updrafts and cold-downdrafts for the same six different stability classes, along with the degree of isotropy. Upon close inspection, we note that these premultiplied PDFs can be divided into two regions which approximately intersect at $(T_{B}\overline{u})/z \approx 1$. The first region extends up to $(T_{B}\overline{u})/z \approx 1$, where the premultiplied PDFs of the warm-updraft and cold-downdraft events collapse with a power-law in highly-convective stability ($-\zeta>$ 2). This power-law region progressively diminishes as the near-neutral stability is approached (figures \ref{fig:14}a to \ref{fig:14}f). The second region extends beyond $(T_{B}\overline{u})/z \approx 1$, where these premultiplied PDFs are widely separated in highly-convective stability, while agreeing with each other in near-neutral stability (figures \ref{fig:14}a to \ref{fig:14}f). In the premultiplied form we can relate the difference in the values between the warm-updrafts and cold-downdrafts to the non-Gaussianity through (\ref{tf_3}). Therefore, we claim that the effect of non-Gaussianity (Gaussianity) in a highly (weakly) convective surface layer is only felt through those warm-updraft and cold-downdraft events having sizes $(T_{B}\overline{u})/z>$ 1. This also explains why at sizes $(T_{B}\overline{u})/z>$ 1, the PDFs and CDFs of the warm-updraft and cold-downdraft events differ most for the highly-convective stability (figures \ref{fig:7}a and \ref{fig:8}a). 

From figure \ref{fig:14} we can also compare the distribution of the degree of isotropy between the warm-updraft and cold-downdraft events. We find that with the change in stability, the peak positions of the degree of isotropy shift systematically from the region $(T_{B}\overline{u})/z<$ 1 to the region $(T_{B}\overline{u})/z>$ 1. From figure \ref{fig:8}, we noted that for sizes $(T_{B}\overline{u})/z>$ 1, the characteristics of the warm-updraft and cold-downdraft events might be related to the passing of the large-scale structures over the measurement points. 

To get a preliminary insight into this systematic shift, we empirically investigated the distributions of the degree of isotropy associated with the warm-updraft and cold-downdraft events by normalizing their streamwise sizes with a mixed length scale. This mixed length scale is a geometric mean of two length scales such as the large-eddy length scale $\lambda$ (see (\ref{model})) and $z$, represented as $z^{0.5}\lambda^{0.5}$. It has been discovered in the context of event based analysis, where \citet{rao1971bursting} showed that the frequency of the burst events in a turbulent boundary layer scaled with a mixed time scale, involving both inner and outer variables. Similarly, \citet{alfredsson1984time} found that the governing time scale of the near-wall region of a channel flow was a mixture of outer and inner scales. They interpreted this as a sign of the interaction of outer and near-wall flows. This mixed scale has been reviewed in details by \citet{buschmann2009near} and \citet{gadeffects}. Recently, \citet{mckeon2017engine} noted that this mixed length scale can be derived from the first principles through matched asymptotic expansions, a theory proposed by \citet{afzal1984mesolayer}. \citet{afzal1982sub,afzal1984mesolayer} showed that by matching the inner and outer expansions of the Reynolds shear stress, an intermediate layer could be formulated for wall-bounded turbulent flows  where the appropriate length scale was the geometric mean of the inner and outer length scales. 

Figure \ref{fig:15} shows that by normalizing the streamwise sizes of the warm-updraft and cold-downdraft events by the mixed length scale could reasonably collapse the peak positions of the degree of isotropy at $(T_{B}\overline{u})/(z^{0.5}\lambda^{0.5}) \approx 0.08$ and $(T_{B}\overline{u})/(z^{0.5}\lambda^{0.5}) \approx 0.15$ respectively. From figures \ref{fig:7} and \ref{fig:14}, we have found that the warm-updraft and cold-downdraft events having sizes $(T_{B}\overline{u})/z<$ 1 are scale-invariant owing to a power-law dependency in the highly-convective stability. This scale-invariant property disappears systematically as the near-neutral stability is approached. Apart from that, the effect of non-Gaussianity (Gaussianity) appears mostly at the sizes $(T_{B}\overline{u})/z>$ 1 in a highly (weakly) convective surface layer. Therefore, this mixed length scaling to collapse the peak positions of the degree of isotropy may suggest that the least anisotropic turbulence might be associated with an interaction between two different physical processes. One of these processes might be related to scale-invariance while the other with non-Gaussianity, associated with the warm-updraft and cold-downdraft events. This is at present a conjecture, which needs to be verified from theoretical arguments. Recently \citet{tong2020velocity} have proposed a matched asymptotic expansion for the convective surface layer, to derive the scaling of the mean velocity profile. By following their footsteps, along with the line of reasoning developed by \citet{afzal1982sub,afzal1984mesolayer}, it might be possible to derive this mixed length scale from the first principles for convective surface layer turbulence. However, this is beyond the scope of the present article. We present our conclusions in the next section.  

\section{Conclusions}\label{conclusion}
We report novel comprehensive results of Reynolds stress anisotropy associated with intermittent heat transport in an unstable ASL, from the SLTEST experimental dataset. We adopt an event-based description of the heat transporting events occurring intermittently and persisting over a wide range of time scales. The Reynolds stress anisotropy is quantified by using a metric called degree of isotropy, computed from the smallest eigenvalue of the anisotropy Reynolds stress tensor. The important results from this study can be broadly summarized as:
\begin{enumerate}
\item The anisotropic state of the Reynolds stress tensor evolves from being dominated by 2-component anisotropy to being dominated by 3-component isotropy as the stability changes from weakly to highly convective. The degree of isotropy of the Reynolds stress tensor is governed by the strength of the vertical velocity fluctuations, which preferentially couple with the temperature fluctuations. These temperature fluctuations exhibit strong (weak) non-Gaussian characteristics in a highly (weakly) convective surface layer. 
\item The Reynolds stress anisotropy in an unstable surface layer is strongly related to the asymmetric and intermittent nature of heat transport, associated with non-Gaussianity in the temperature fluctuations. 
\item By adopting an event based approach, it is found that not all the heat flux events are associated with same anisotropic state of turbulence. The anisotropic states associated with highly-intermittent large heat flux events are dominated by 1-component anisotropy. Whereas, the anisotropic states associated with more frequent but weak heat flux events are dominated by 2-component anisotropy. On the other hand, the anisotropic states associated with moderate heat flux events which lie between these two extremes are dominated by 3-component isotropy.  
\item There is a critical size associated with the organized heat flux events (warm-updrafts and cold-downdrafts) which corresponds to the maximum value of the degree of isotropy (i.e. least anisotropic turbulence). By investigating the anisotropic states of the Reynolds stress tensor, it is found that in a highly-convective surface layer, the warm-updraft and cold-downdraft events smaller (larger) than this critical size, are associated with anisotropic states dominated by 2-component (1-component) anisotropy. However, in a near-neutral surface layer the anisotropic states are mostly dominated by 2-component anisotropy, regardless of the sizes of the warm-updraft and cold-downdraft events.
\item This critical size associated with least anisotropic turbulence does not scale with $z$. However, the $z$-scaling is successful in collapsing the peak positions of the heat flux distribution associated with the sizes of the warm-updraft and cold-downdraft events. This disagreement occurs because the sizes of the warm-updraft events corresponding to maximum heat flux are also associated with significant amount of streamwise momentum. This causes a drop in the degree of isotropy associated with their sizes. 
\end{enumerate}

Note that the findings from this study should be verified from the field experiments in an unstable ASL flow conducted over the rough surfaces and in complex terrains. Our preliminary investigation shows that this critical size probably scales with a mixed-length scale $z^{0.5}\lambda^{0.5}$, where $\lambda$ is the large-eddy length scale. We propose a conjecture that this mixed-length scaling may reflect an interaction between two different physical processes, one of which may be associated with scale-invariance and the other with the non-Gaussianity in turbulence. The verification of this conjecture is beyond the scope of the present study. An inevitable limitation of this study is the unavailability of three-dimensional velocity and temperature information. Due to this constraint, the intermittent heat flux events and the associated Reynolds stress anisotropy, cannot be connected to the three-dimensional topology of the coherent structures in convective turbulence. In the future, we would address this problem through large eddy or direct numerical simulations. This study also raise a few important questions which deserve future attention: 
\begin{enumerate}
\item Is there a theoretical framework to explain the mixed length scale in convective turbulence?
\item What is the physical connection between the event based (related to flow structures) and scale based (related to harmonic analysis) description of turbulence anisotropy? 
\end{enumerate}

\section*{Acknowledgements}
Indian Institute of Tropical Meteorology (IITM) is an autonomous institute fully funded by the Ministry of Earth Sciences. The author Tirtha Banerjee acknowledges the new faculty start-up funding from the Department of Civil and Environmental Engineering, the Henry Samueli School of Engineering, University of California, Irvine. The authors would like to thank the anonymous reviewers whose comments helped to improve the quality of the article. The authors are also grateful to Keith G McNaughton for letting them use the SLTEST dataset for this research and pointing out some references on mixed length scaling. The computer codes used in this study are available to all the researchers by contacting the corresponding author.  
\clearpage

\appendix
\section{Histograms of the heat flux events}\label{appA}
\begin{figure*}
\centering
\vspace*{0.5in}
\hspace*{-0.85in}
\includegraphics[width=1.3\textwidth]{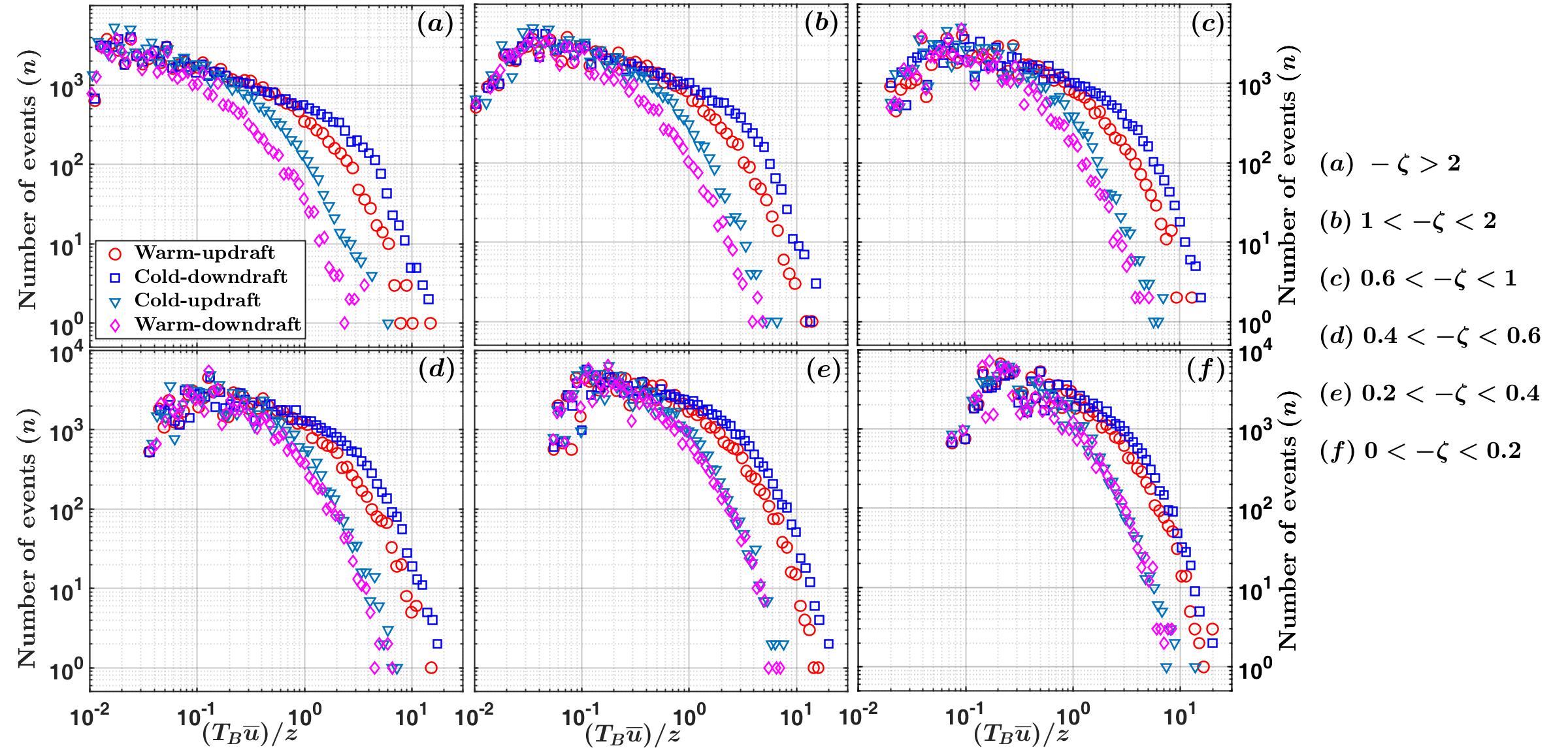}
  \caption{Same as in figure \ref{fig:7}, but the histograms are shown for the heat flux events from each quadrant.}
\label{fig:16}
\end{figure*}

In figures \ref{fig:16}a--f, we show the histograms of the heat flux events from each quadrant of $T^{\prime}$-$w^{\prime}$, corresponding to the six different stability classes. The number of events ($n$) shown in figures \ref{fig:16}a--f are computed after considering all the 30-min runs from a particular stability class (e.g., 55 number of 30-min runs for $-\zeta>$ 2, amounting to 27.5 hours of observation). For each stability class, the total number of heat flux events counted over all the sizes $(T_{B}\overline{u})/z$ from each quadrant are given in table \ref{tab:4}. Typically, for the warm-updraft and cold-downdraft quadrants we encounter more than 100-200 number of heat flux events corresponding to the sizes $(T_{B}\overline{u})/z>4$. For the counter-gradient quadrants the total number of heat flux events corresponding to large sizes $(T_{B}\overline{u})/z>1$ is also more than 100, although the histograms decrease faster than the down-gradient quadrants. This implies that these counter-gradient events have a statistical tendency to occur in smaller sizes and do not persist for a long time. Therefore, the mean statistics shown in figures \ref{fig:7}--\ref{fig:15} for the heat flux events from all the four quadrants have been averaged over more than 100-200 number of events for the streamwise sizes $(T_{B}\overline{u})/z>1$. We note that many statistics textbooks \citep[e.g.,][]{ross2014introduction} as well as the seminal paper by \citet{student1908probable} consider that a sample size of more than 30 is enough for ensuring the statistical convergence of the mean to the actual population mean, from the weak law of large numbers. Nevertheless, we also performed the Student's t-test to ensure the statistical significance of the ensemble mean. For the large values of $(T_{B}\overline{u})/z$, based on the sample size of around 100-200 events, the margin of error in the ensemble mean computed over these samples is about 7-10\% with a confidence level of 95\%. 

\begin{table}
  \begin{center}
\def~{\hphantom{0}}
  \begin{tabular}{lcccc}
Stability class (\#) & Warm-updraft & Cold-downdraft & Cold-updraft & Warm-downdraft
\\
\\
$-\zeta \ >$ 2 (55) & 63608 & 66482 & 71547 & 54523\\
1 $< \ -\zeta \ <$ 2 (53) & 76621 & 80812 & 83334 & 66543\\
0.6 $< \ -\zeta \ <$ 1 (41) & 70043 & 73814 & 73472 & 60849\\
0.4 $< \ -\zeta \ <$ 0.6 (34) & 70062 & 72961 & 69641 & 61341\\
0.2 $< \ -\zeta \ <$ 0.4 (44) & 105273 & 108822 & 101299 & 91870\\
0 $< \ -\zeta \ <$ 0.2 (34) & 100780 & 103151 & 92335 & 88357\\
  \end{tabular}
  \caption{The total number of heat flux events from each quadrant of $T^{\prime}$-$w^{\prime}$ are tabulated for all the sizes $T_{B}\overline{u}/z$, corresponding to each stability class as shown in table \ref{tab:2}. The symbol \# denotes the number of 30-min runs in each stability category.}
  \label{tab:4}
  \end{center}
\end{table}

\section*{Supplementary material}
This paper has supplementary figures shown at the end.

\section*{Declaration of Interests} 
The authors report no conflict of interest.
\clearpage

\bibliographystyle{jfm}
\bibliography{jfm-instructions}

\clearpage
\renewcommand{\thefigure}{S\arabic{figure}}
\setcounter{figure}{0}

\begin{figure*}
\centering
\vspace*{0.5in}
\hspace*{-2.5in}
\includegraphics[width=1.8\textwidth]{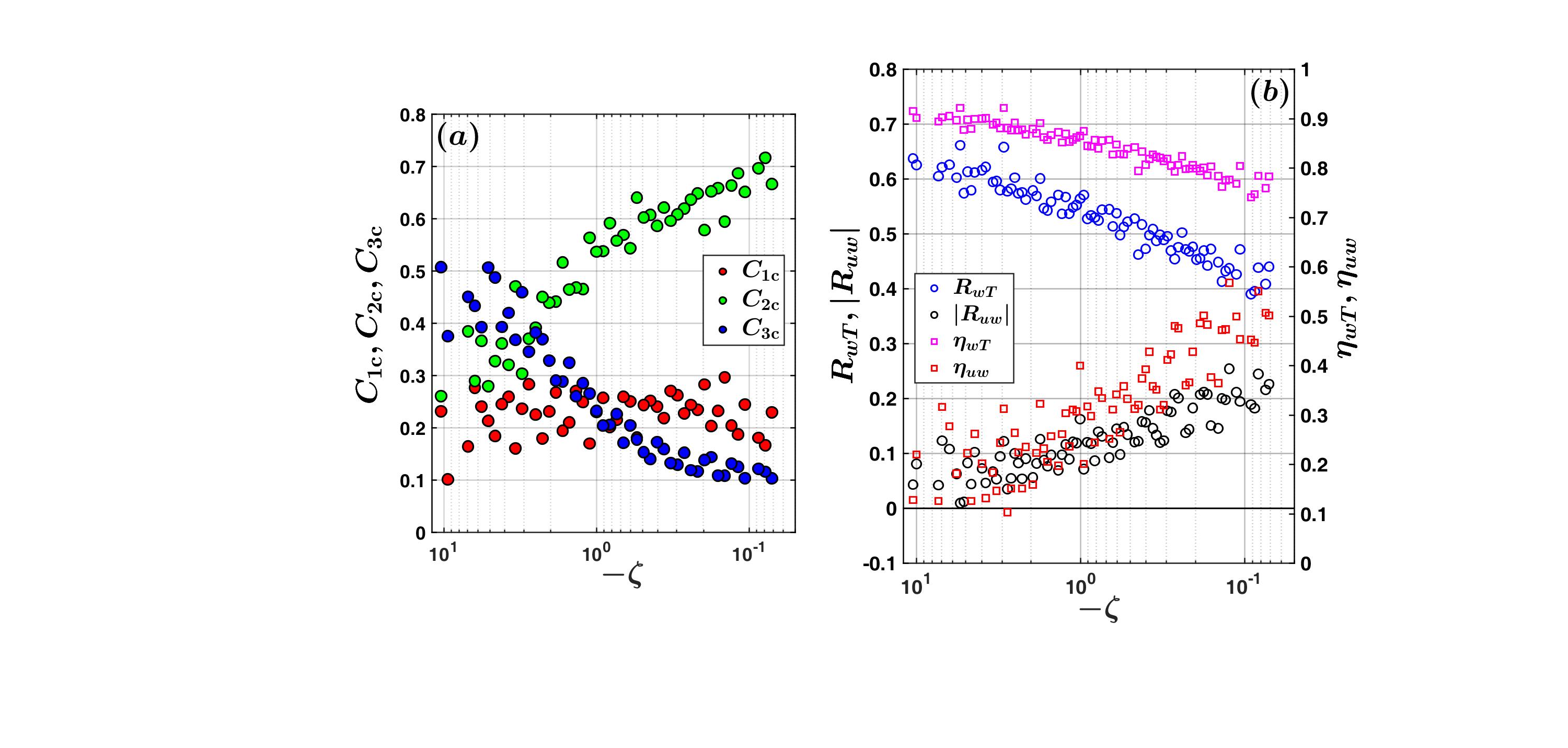}
  \caption{The (a) three anisotropy coefficients ($C_{1 \rm c}$, $C_{2 \rm c}$ and $C_{3 \rm c}$) and (b) correlation coefficients between $w$ and $x$ ($R_{wx}$, where $x$ can be $T$ or $u$) along with the transport efficiencies of heat and momentum ($\eta_{wT}$, $\eta_{uw}$) are plotted against the stability ratio $-\zeta$. The correlation coefficients are shown on the left $y$ axis of panel (b), whereas the transport efficiencies are shown on the right $y$ axis. Note that the absolute values of $R_{uw}$ ($|R_{uw}|$) are plotted instead of their original negative values and the $x$ axis is reversed such that the $-\zeta$ values proceed from large to small.}
\label{fig:s1}
\end{figure*}

\begin{figure}
\centering
\vspace*{0.5in}
\hspace*{-0.75in}
\includegraphics[width=1.25\textwidth]{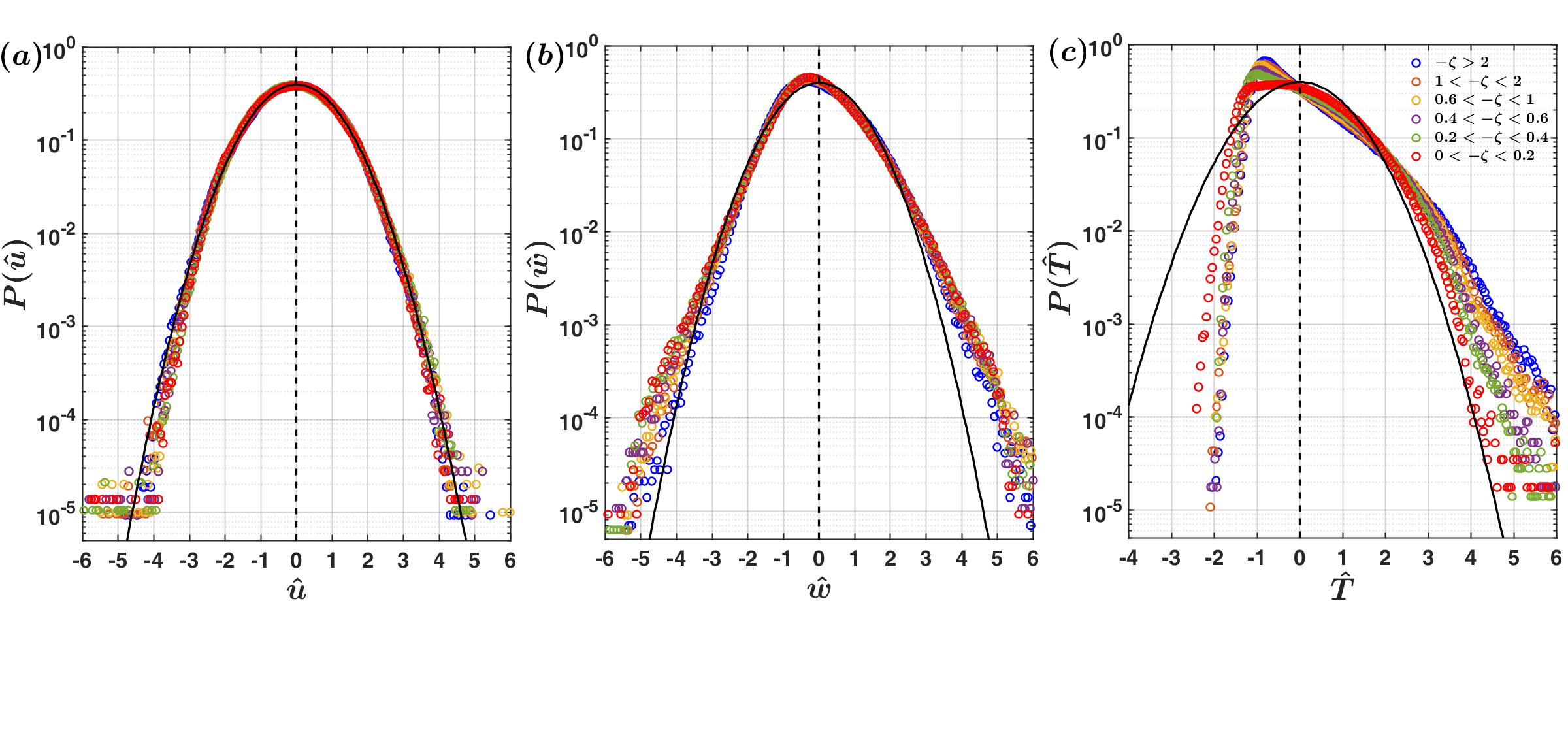}
 \caption{The PDFs of (a) $u^{\prime}/\sigma_{u}$ ($\hat{u}$), (b) $w^{\prime}/\sigma_{w}$ ($\hat{w}$), and (c) $T^{\prime}/\sigma_{T}$ ($\hat{T}$) are shown for the six different classes of the stability ratio as indicated in the legend on panel (c). The thick black lines on all the panels represent the Gaussian distribution.}
\label{fig:s2}
\end{figure}

\begin{figure}
\centering
\vspace*{0.5in}
\hspace*{-0.75in}
\includegraphics[width=1.25\textwidth]{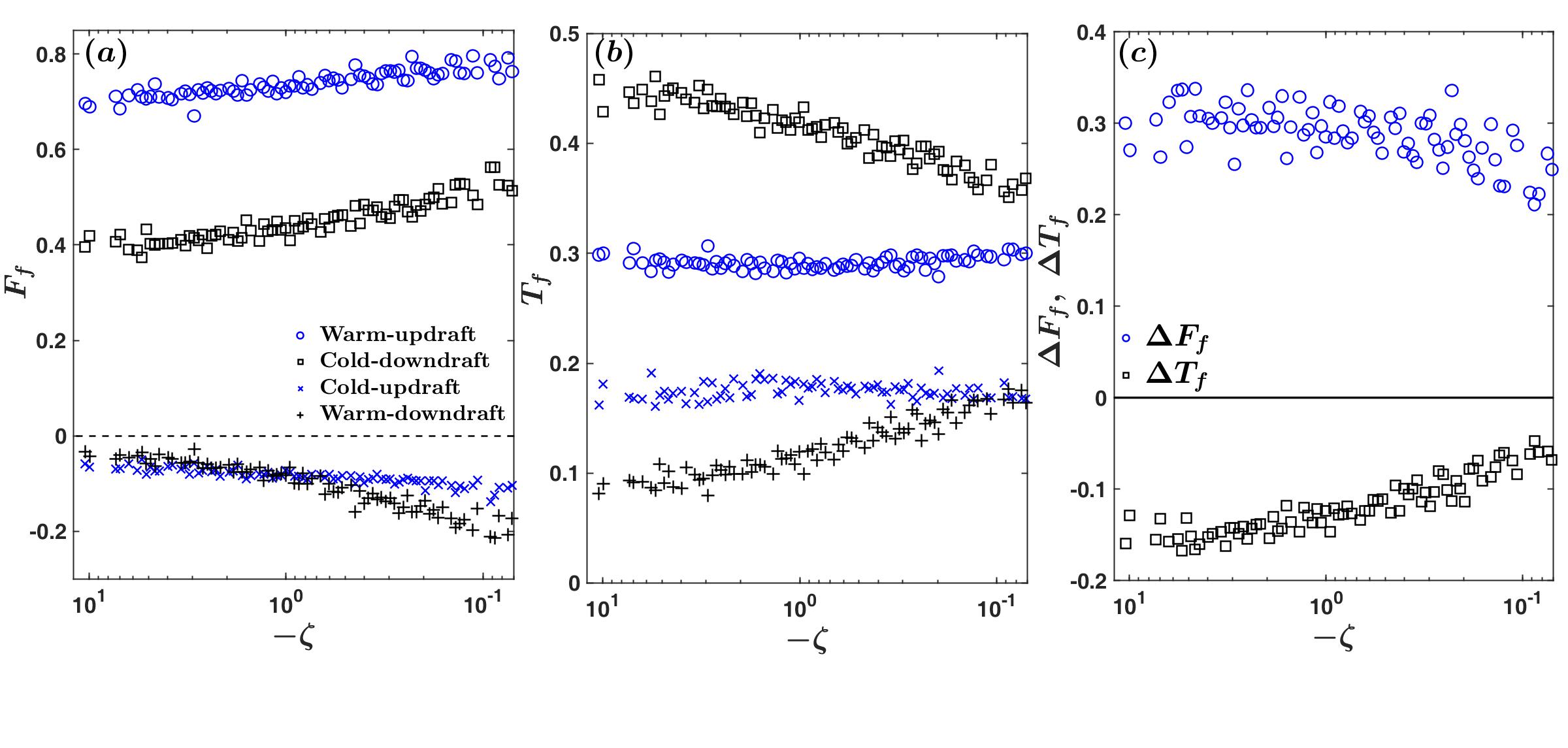}
 \caption{The (a) heat flux fractions ($F_{f}$), and (b) time fractions ($T_{f}$) associated with the four different quadrants of $T^{\prime}$-$w^{\prime}$ plane, as indicated in the legend in panel (a). The differences in flux fractions ($\Delta F_{f}$) and time fractions ($\Delta T_{f}$) between the warm-updraft and cold-downdraft quadrants are shown in panel (c).}
\label{fig:s3}
\end{figure}

\end{document}